\documentclass{tac}

\author {Krzysztof Worytkiewicz}

\thanks{
The author wishes to thank the anonymous referee for the constructive part of 
his criticism. He gratefully acknowledges Ronnie Brown, Eric Goubault and 
Vincent Schmitt for the stimulating discussions. His debt to Claudio Hermida 
deserves a separate sentence.
}

\address{Ollon, Switzerland\\
}

\title {Synchronization from a Categorical Perspective}

\copyrightyear{2004}

\eaddress{krisw@bluewin.ch}

\usepackage{amsmath}
\usepackage{amssymb}

\usepackage{latexsym}

\usepackage[2cell,v2]{xy}
\UseTwocells
\UseHalfTwocells

\newcommand{\leirom}{\renewcommand{\labelenumi}{\textit{(\roman{enumi})}}}
\newcommand{\leialph}{\renewcommand{\labelenumi}{(\alph{enumi})}}
\newcommand{\leiAlph}{\renewcommand{\labelenumi}{(\Alph{enumi})}}
\newcommand{\leiarab}{\renewcommand{\labelenumi}{(\arabic{enumi})}}

\newcommand{\leiialph}{\renewcommand{\labelenumii}{(\alph{enumii})}}
\newcommand{\leiiAlph}{\renewcommand{\labelenumii}{(\Alph{enumii})}}
\newcommand{\leiiarab}{\renewcommand{\labelenumii}{(\arabic{enumii})}}

\newcommand{\leiiialph}{\renewcommand{\labelenumii}{(\alph{enumiii})}}
\newcommand{\leiiiAlph}{\renewcommand{\labelenumii}{(\Alph{enumiii})}}
\newcommand{\leiiiarab}{\renewcommand{\labelenumii}{(\arabic{enumiii})}}

\renewcommand{\emptyset}{\varnothing}

\newcommand{\N}{\mathbb{N}}

\newcommand{\free}{\mathbf{\mathcal{F}}}

\newcommand{\one}{\mathrm{id}}

\newcommand{\onee}{1}

\newcommand{\tmname}[1]{\textsc{#1}}
\newcommand{\tmem}[1]{{\em #1\/}}
\newcommand{\tmmathbf}[1]{\mathbf{#1}}
\newcommand{\tmop}[1]{{#1}}

\newtheorem{definition}{Definition}
\newtheorem{proposition}{Proposition}
\newcommand{\dueto}[1]{\textup{{(#1) }}}

\newenvironment{enumerateroman}
{\leirom \begin{enumerate}}
{\end{enumerate} \leiarab}

\newenvironment{itemizeminus}
  {\begin{itemize}}{\end{itemize}}

\newenvironment{enumeratenumeric}{\begin{enumerate}}{\end{enumerate}}

\newenvironment{theproof}{
  \noindent\textsc{Proof.}\ }{\hspace*{\fill}
\begin{math}
\Box
\end{math}
\medskip}

\newcommand{\equallim}{\mathop{=}\limits}
\newcommand{\tmstrong}[1]{\textbf{#1}}
\newcommand{\tmscript}[1]{\text{\scriptsize $#1$}}
\newtheorem{theorem}{Theorem}
\newtheorem{varremark}{Remark}

\newtheorem{lemma}{Lemma}

\newcommand{\longrightarrowlim}{\mathop{\longrightarrow}\limits}
\newcommand{\longleftarrowlim}{\mathop{\longleftarrow}\limits}

\newtheorem{corollary}{Corollary}
\newcommand{\Longleftrightarrowlim}{\mathop{\Longleftrightarrow}\limits}
\newcommand{\rightarrowlim}{\mathop{\rightarrow}\limits}

\newcommand{\deq}{{\equallim}^{def.}}
\newcommand{\idle}{\star}

\newcommand{\nin}{\not\in}

\newcommand{\mathbbm}[1]{\mathbb{#1}}

\newcommand{\boxast}{\otimes}

\newcommand{\assign}{:=}

\newcommand{\llbracket}{[\![}
\newcommand{\rrbracket}{]\!]}

\newcommand{\paragraf}[1]{\mbox{}\\\hspace{.25em}\textit{#1}.}

\begin{document}

\maketitle
\begin{abstract}
We introduce a notion of synchronization for higher-dimensional automata,
  based on coskeletons of cubical sets. Categorification transports this
  notion to the setting of categorical transition systems. We apply the
  results to study the semantics of an imperative programming language with
  message-passing.
\end{abstract}

{\tableofcontents}


\section{Introduction}

Traditional labeled transition systems synchronize on labels
{\cite{NielsenM:modc1}}, essentially by a limit construction. Any type of
synchronization of transition systems can be represented that way. The
situation for higher-dimensional automata or HDA's
{\cite{eric:cmcim}} is not so straightforward. Indeed, in the one-dimensional
case a transition is either in the limiting object or is not. On the other
hand, in the case of general  HDA's a transition has a {\tmem{dimension}} and
synchronization may induce a change of dimension in addition to filter the
transition out. Since HDA's live in slices of the category
$\tmmathbf{\tmop{cSet}} $ of cubical sets, there is a well-established theory
available. In this paper, the relevant ingredients are the 
$n$-{\tmem{coskeleton}} (right adjoint to the
$n$-{\tmem{truncation}}) and the {\tmem{categorification}} (left adjoint 
to the {\tmem{cubical
nerve}}). We introduce a notion of $(\Sigma, n)$-coskeleton, a version of the
usual $n$-coskeleton which interacts well with the (higher-dimensional)
labeling. It turns out that this yields a notion of synchronization of HDA's
suitable for the study of message passing.

Section \ref{sec-gircond} contains some categorical background. In section
\ref{sec-cubes}, we recall the salient facts about cubical sets and prove some
relevant technical lemmas. In section  \ref{sec-hdasync}, we recall what HDA's
are and introduce their 1-{\tmem{coskeletal synchronization}}. In a nutshell,
all potential boundaries made of unsynchronized transitions are
filled. This applies in particular to transition systems, since the latter
are nothing but 1-dimensional HDA's. A transition system is {\tmem{acubic}} if
it does not contain boundaries, i.e. paths ``around'' cubes. Our main result
about 1-coskeletal synchronization is theorem \ref{theo-catsynchro}. Without
technical noise, this theorem can be stated as follows:
\\
\\
{\bf Theorem} {\em The categorification of 1-coskeletally synchronized acubic
 transition systems is a limit of free categories.}
\\
\\
{\noindent}After having introduced the programming language {\tmname{CIP}} in
section \ref{sec-cip}, we study its semantics in section \ref{sec-evo}. The
categorification of HDA's obtained by applying {\tmname{CIP}}'s operational
rules yields {\tmem{categories of evolutions}} and {\tmem{control
categories}}, technically domains and codomains of ulf functors. The
Giraud-Conduch\'e correspondence of section \ref{sec-gircond} yields a
finite presentation in form of {\tmem{categorical transition systems}},
technically pseudofunctors from free categories to $\tmmathbf{\tmop{Span}}$.
Section \ref{sec-sim} takes up the topic of  (bi)simulation and {\tmem{process
categories}}. Concluding remarks are to be found in section \ref{sec-conc}.

\section{\label{sec-gircond}The Giraud-Conduch\'e Correspondence}

In this section, we expose some known and less known facts about functors with
the {\tmem{unique lifting of factorizations}} property
{\cite{street-cat-struct}} and their correspondence with certain
pseudofunctors. For general background about bicategories, we refer to the
original text {\cite{bena}} and to {\cite{bor1}}.

\subsection{\label{sec-2-ulf}Ulf Functors}

\begin{definition}
  A functor $F : \mathbb{B} \rightarrow \mathbb{C}$ has the \emph{unique
  lifting of factorizations} property if, given $u \in \mathbb{B}$
  and $\mathbb{C} \ni f = h \circ g$ such that $F \left( u \right) = f$, there
  are unique $v, w \in \mathbb{B}$ such that $u = w \circ v$ with $F \left( v
  \right) = g$ and $F \left( w \right) = h$.
\end{definition}

We use the acronym {\tmem{ulf}} for such functors.

\begin{proposition}
  \label{prop-streetulf} 
\dueto{{Street {\cite{street-cat-struct}}}}
Let $F :
  \mathbb{B} \rightarrow \mathbb{C}$ be a functor, $J = J_2$ be the interval
  category
  \[ - \longrightarrow + \]
  $J_3$ be the category generated by
  \[ - \longrightarrow 0 \longrightarrow + \]
  and $d_1 : J_2 \longrightarrow J_3$ be the functor such that $- \mapsto -$ and
  $+ \mapsto +$. The following are equivalent
  \begin{enumerateroman}
    \item $F$ is an ulf functor;
    
    \item $F$ has the strict right lifting property with respect to $d_1$, i.e.
    any commuting square

    \begin{center}
      $
\xymatrix {I_2 \ar[r] \ar[d]_{d_1} & \mathbb{B} \ar[d]^{F} 
\\
I_3 \ar@{.>}[ru] \ar[r] & \mathbb{C} 
}
$

    \end{center}
    
    {\noindent}admits a unique filler.
  \end{enumerateroman}
\end{proposition}

\begin{theproof}
  Obvious.
\end{theproof}

Item $( ii )$ of Proposition \ref{prop-streetulf} is actually taken as
definition in \cite[pp. 532-533]{street-cat-struct}.

\begin{proposition}
  \label{prop-streetfree} {\dueto{Street {\cite{street-cat-struct}}}}Let $(
  \mathbb{N}, + )$ be the additive monoid on the natural numbers. A category
  $\mathbb{A}$ is free iff there is an ulf
  functor $\ell : \mathbb{A} \longrightarrow ( \mathbb{N}, + )$.
\end{proposition}

\begin{theproof}
  Obvious.
\end{theproof}


\begin{proposition}
  \label{prop-ulfcomp}The full subcategory
  $\tmmathbf{Ulf}^{\rightarrow} \subseteq
  \tmmathbf{\tmop{Cat}}^{\rightarrow}$ with ulf functors as objects is
  complete.
\end{proposition}

\begin{theproof}
  Products are straightforward. Suppose

  \begin{center}
    $
\xymatrix {\mathbb{E}\; \ar@{^{(}->}[r] \ar[dd]_{K} &
\mathbb{A} \ar[dd]_U \rtwocell^F_G{\omit} & \mathbb{B} \ar[dd]^V
\\
&
\\
\mathbb{E}' \; \ar@{^{(}->}[r] & \mathbb{A}' \rtwocell^{F'}_{G'}{\omit} 
& \mathbb{B}'
}
$

  \end{center}

  {\noindent}is an equalizer diagram in $\tmmathbf{\tmop{Cat}}^{\rightarrow}$,
  so both rows are equalizer diagrams in $\tmmathbf{\tmop{Cat}}$ and $K =
  U|_{\mathbb{A}}$. Suppose $U$ and $V$ are ulf functors. Let $f \in
  \mathbb{E}$ and suppose $K ( f ) = f'$ and $f' = q' \circ p'$ is a
  factorization in $\mathbb{E}'$. Since $U$ is an ulf functor, there is a
  unique factorization $f = q \circ p$ in $\mathbb{A}$ such that $U ( p ) =
  p'$ and $U ( q ) = q'$. Let $f_1 \equallim^{\tmop{def} .} F ( f )$. Since $f
  \in \mathbb{E}$, we have $f_1 = F ( f ) = G ( f )$. Let $f_1'
  \equallim^{\tmop{def} .} V ( f_1 )$, $p_1' \equallim^{\tmop{def} .} F' ( p'
  )$ and $q_1' \equallim^{\tmop{def} .} F' ( q' )$. Since $p' \in
  \mathbb{E}'$ and $q' \in \mathbb{E}'$, the factorization
  \[ f_1' = q_1' \circ p_1' \]
  has liftings $f_1 = F ( q ) \circ F ( p )$ and $f_1 = G ( q ) \circ G ( p
  )$. But $V$ is an ulf functor, hence $F ( p ) = G ( p )$ and $F ( q ) = G (
  q )$, i.e. the unique lifting of the factorization $f = q \circ p$ is in
  $\mathbb{E}$.
\end{theproof}

\subsection{Lax Comma Categories}

\begin{definition}
  Let $\mathcal{K}$ be a bicategory. A morphism in $\mathcal{K}$ is a
  \textit{map} if it admits a right adjoint.
\end{definition}

\begin{proposition}
  Let $\mathbb{B}$ be a category with pullbacks. A morphism in
  $\tmmathbf{\tmop{Span}} ( \mathbb{B} )$ is a map precisely when its left
  leg is iso.
\end{proposition}

It follows that the iso class of a map in $\text{$\tmmathbf{\tmop{Span}} (
\mathbb{B} )$}$ has a span with identity left leg as a canonical
representant.

Let $F, G : \hspace{0.25em} \mathcal{K} \rightarrow \mathcal{L}$ be lax
functors. Recall that a lax transformation
\[ \alpha : \hspace{0.25em} F \Rightarrow G \]
is given by the data
\begin{enumerateroman}
  \item for each $x \in \mathcal{K}$ a morphism $\mathcal{L} \ni \alpha_x :
  \hspace{0.25em} F \left( x \right) \rightarrow G \left( x \right)$;
  
  \item for each morphism $\mathcal{K} \ni f : \hspace{0.25em} x \rightarrow
  y$ a 2-cell

  \begin{center}
    \begin{center}
      $\xymatrix{
F(x) \ar[r]^{\alpha_x} \ar[d]_{F(f)} 
& G(x) \ar[d]^{G(f)} 
\\
F(y) \ar[r]_{\alpha_y} & 
G(y) 
\ultwocell<\omit>{\alpha_f}
}$

    \end{center}
  \end{center}

\end{enumerateroman}
subject to the coherence conditions
\begin{enumerateroman}
  \item $\alpha_{f'} \bullet \left( G \left( \theta \right) \circ \alpha_x
  \right) = \alpha_f$ for each 2-cell $\theta : \hspace{0.25em} f \Rightarrow
  f' : x \rightarrow y$;
  
  \item $\left( \alpha_g \circ F \left( f \right) \right) \bullet \alpha_f =
  \alpha_{g \circ f}$ for each $f : \hspace{0.25em} x \rightarrow y$ and $g :
  \hspace{0.25em} y \rightarrow z$.
\end{enumerateroman}
\begin{definition}
  Let $F, G : \hspace{0.25em} \mathcal{K} \rightarrow \mathcal{L}$ be lax
  functors. A lax transformation is \textit{representable} provided its
  \textit{components} $\alpha_x$ are maps for all $x \in \mathcal{K}$.
\end{definition}

Special cases of interest are lax representable transformations among lax
functors from a category to $\tmmathbf{\tmop{Span}} \equallim^{\tmop{def} .} 
\tmmathbf{\tmop{Span}} ( \tmmathbf{\tmop{Set}} )^{}$. Let $p, q :
\hspace{0.25em} \mathbb{B} \rightarrow \tmmathbf{\tmop{Span}}$ be such lax
functors and let $\alpha : \hspace{0.25em} p \Rightarrow q$ be a lax
representable transformation. Its data with respect to to $\mathbb{B} \ni f :
\hspace{0.25em} x \rightarrow y$ is given by the lax square

\begin{center}
  $
\xymatrix @=5mm{
        p(x) & 
	p(x) \ar[l]_{\one} \ar[r]^{\alpha_x} \xtwocell[1,1]{}\omit{^<6>{\alpha_f}} & 
	q(x)
	\\
	p(f) \ar[u]^{p(f)_1} \ar[d]_{p(f)_2} &
	&
	q(f) \ar[u]_{q(f)_1} \ar[d]^{q(f)_2}
	\\
	p(y)& 
	p(y) \ar[l]^{\one} \ar[r]_{\alpha_y} & 
	q(y)
}
$
\end{center}

{\noindent}i.e. $\alpha_f$ is a morphism of spans

\begin{center}
  $
\xymatrix @=5mm{
	&
	p(f) \ar[dl]_{p(f)_1} \ar[dr]^{\alpha_y\:\circ\:p(f)_2} \ar[dd]_{\alpha_f} 
	&
	\\
	p(x) 
	&
	&
	q(y)
	\\
	&
	p(x)\:\times_{q(x)}\:q(f) \ar[ul]_{p_1} \ar[ur]^{q(f)_2\:\circ\:p_2}
	&
}
$
\end{center}

Recall that a lax functor $\text{$F : \hspace{0.25em} \mathcal{K} \rightarrow
\mathcal{L}$}$  is {\tmem{normalized}} or \tmem{normal} provided it is strict on identities,
i.e. the distinguished 2-cell $\tmop{id}_{F ( x )} \Rightarrow F ( \tmop{id}_x
)$ is the identity 2-cell for all $x \in \mathcal{K}$.

\begin{definition}
  \label{def-superlax} Let $\tmmathbf{\tmop{rGraph}}$ be the category of
  reflexive graphs, i.e. graphs with a distinguished loop at every vertex. Let
  $\mathcal{K}$ be a bicategory and $\mathcal{F} : \tmmathbf{\tmop{rGraph}}
  \longrightarrow \tmmathbf{\tmop{Cat}}$ the ''free'' functor. The lax
  comma-category $\mathcal{F} /\!/ \mathcal{K}$ is
  given by
  \begin{enumerateroman}
    \item Objects: normal pseudofunctors from a path category to
    $\mathcal{K}$;
    
    \item Morphisms: given $s : \hspace{0.25em} \mathcal{F (} G ) \rightarrow
    \mathcal{K}$ and $t : \hspace{0.25em} \mathcal{F (} H ) \rightarrow
    \mathcal{K}$ normal pseudofunctors, a morphism $\alpha :
    \hspace{0.25em} s \rightarrow t$ is a lax representable transformation
    $\alpha : \hspace{0.25em} s \Rightarrow t \circ \mathcal{F} ( k )$ for
    a morphism of reflexive graphs $k : \hspace{0.25em} G \rightarrow H$ ;
    
    \item Composition: $\beta \circ \alpha = \left( l \circ k, l \beta \circ
    \alpha \right)$ where $\alpha : \hspace{0.25em} s \Rightarrow t \circ
    \mathcal{F} ( k )$ and $\beta : \hspace{0.25em} t \Rightarrow u \circ
    \mathcal{F} ( l )$ while the vertical composition $l \beta \circ \alpha$
    is given by componentwise bicategorical pasting.
  \end{enumerateroman}
\end{definition}

\subsection{\label{sec-2-thecorr}The Correspondence}

Let $\mathbf{\tmop{Ulf}} / \mathbb{B}$ be the full subcategory of the slice
category $\mathbf{\tmop{Cat}} / \mathbb{B}$ where the objects are ulf
functors. Let $\text{{{Psd}}} \left[ \mathbb{B},
\text{{{{\tmstrong{Span}}}}} \right]$ be the category of normal
pseudo-functors from $\mathbb{B}$ to {{{\tmstrong{Span}}}} and lax
representable transformations. There is the equivalence of categories
\[ \mathbf{\tmop{Ulf}} / \mathbb{B} \hspace{0.5em} \simeq \hspace{0.5em}
   \text{{{Psd}}} \left[ \mathbb{B}, \text{{{{\tmstrong{Span}}}}}
   \right] \]
which is the discrete particular case of the equivalence between functors with
liftings of factorizations unique up to connected component and distributors,
discovered independently by Giraud and by Conduch\'e in the early 70's
{\cite{JohnstonePT:dicf}}.

\begin{proposition}
  \label{prop-ulf}Let $\mathbf{\tmop{Ulf}}_{\mathcal{F}}^{\hspace{0.25em}
  \rightarrow}$ be the full subcategory of the comma-category
  $\tmop{id}_{\mathbf{\tmop{Cat}}} \hspace{0.25em} \downarrow \hspace{0.25em}
  \mathcal{F}$ where the objects are ulf functors and
  \[ \begin{array}{l}
       \;\; \mathbb{A}\\
       {\scriptstyle u} \downarrow\\
       \mathcal{F} H
     \end{array} \in \mathbf{\tmop{Ulf}}_{\mathcal{F}}^{\hspace{0.25em}
     \rightarrow} \]
  Then there is a reflexive graph $G$ and a morphism $m : G \longrightarrow H$
  such that $u = \mathcal{F} ( m )$.
\end{proposition}

\begin{theproof}
  It is obvious that the composition of two ulf functors is an ulf functor. By
  proposition \ref{prop-streetfree} there is an ulf functor $\ell :
  \mathcal{F} ( H ) \longrightarrow ( \mathbb{N}, + )$. Hence $\ell \circ u$
  is ulf so $\mathbb{A}$ is free again by proposition \ref{prop-streetfree}.
  But $u$ is ulf so it preserves length. In particular, generators are mapped
  on generators.
\end{theproof}

A variation of the classic equivalence above is

\begin{theorem}
  \label{theo-gircond}Let $\mathbf{\tmop{Ulf}}_{\mathcal{F}}^{\hspace{0.25em}
  \rightarrow}$ be as in proposition \ref{prop-ulf}. There is an equivalence
  of categories
  \[ \mathbf{\tmop{Ulf}}_{\mathcal{\mathcal{F}}}^{\hspace{0.25em} \rightarrow}
     \simeq \mathcal{F} /\!/ \tmmathbf{\tmop{Span}} \]
\end{theorem}

\begin{theproof}
  The ulf counterpart $\pi_s : \overline{s} \rightarrow \mathcal{F} ( G )$ of
  a pseudofunctor $s : \mathcal{F} ( G ) \rightarrow \tmmathbf{\tmop{Span}}$
  is obtained by an appropriate Grothendieck construction: a morphism in
  $\overline{s}$ is of the form $\left( k, f \right) : \hspace{0.25em} \left(
  a, x \right) \rightarrow \left( b, y \right)$ with $a \in s \left( x
  \right)$ and $b \in t \left( x \right)$ while $k \in s \left( f \right)$
  such that $k$ it is mapped on $a$ respectively $b$ by $s \left( f \right)$'s
  left respectively right leg. We further have

  \begin{center}
    $\xymatrix@=15mm{  
\bar{s} \ar @{}[dr] |{(*)} \ar[d]_{\pi_s} \ar[rr]^{\bar{\alpha}} && \bar{t} \ar[d]^{\pi_t} 
\\
\mathcal{F}(G) \ar[dr]_s \ar[rr]^{\mathcal{F}(h)} && \mathcal{F}(H) \ar[dl]^t
\dtwocell\omit{^<13>\alpha}
\\
& \mathbf{Span} &
}$

  \end{center}

  {\noindent}where

  \begin{center}
    $\xymatrix@=3mm{  
& (a,x) \ar[dd]^{(k,f)} & & &
(\alpha_x(a),\free (h) (x) ) \ar[dd]^{((p_2 \: \circ \: \alpha_f)(k),\free (h)(f))}
\\
\bar{\alpha} : & & & \mapsto &
\\
& (b,y) & & & (\alpha_y(b),\free (h)(y))
}$

  \end{center}

  The other way round it is enough to take the fibers. Let $s :
  \hspace{0.25em} \mathbb{D} \rightarrow \mathcal{F} ( G )$ be an ulf functor
  and let $s_x$ respectively $s_f$ be the set of objects over $\left. x \in
  \mathcal{F} ( G \right)_0$ respectively the set of arrows over $\left. f \in
  \mathcal{F} ( G \right)_1$. The functor $s$ determines a pseudofunctor
 $\underline{s}
  : \mathcal{F} (G) \rightarrow \tmmathbf{\tmop{Span}}$ given by

  \begin{center}
    $\xymatrix@=5mm{  
& x \ar[dd]^f & & & s_x
\\
\underline{s} : & & & \mapsto & s_f \ar[u]_{dom} \ar[d]^{cod}
\\
& y & & & s_y
}$
  \end{center}

  Let $\left( m, n \right) : \hspace{0.25em} s \rightarrow t$ be a morphism in
  $\mathbf{\tmop{Ulf}}_{\mathcal{F}}^{\hspace{0.25em} \rightarrow}$. It
  determines a lax representable transformation $\underline{m} :
  \hspace{0.25em} \underline{s} \Rightarrow n \circ \underline{t}$ given at
  $f$ by $\underline{m}_{\: x} \equallim^{\tmop{def} .} m \mid_{s_x}$ and
  $\underline{m}_{\:f} \left( p \right) \equallim^{\tmop{def} .} \left(
  \text{{\tmem{dom}}} (p), m \left( p \right) \right)$.
\end{theproof}

\begin{remark}
  \label{rem-ulfarrows}The square $\left( \ast \right)$ is a pullback square
  provided $\alpha = \tmop{id}$.
\end{remark}

\begin{remark}
  $\mathcal{F} /\!/ \mathbf{Span}$ is complete by theorem
  \ref{theo-gircond} and proposition \ref{prop-ulfcomp}.
\end{remark}

We indicate for reference the construction of pullbacks at hand of the case
where the involved pseudo-functors have the same domain $\mathcal{\mathcal{F}
(} K )$. No generality is lost doing so since it is always possible to
reindex. Let thus $F, G, H : \hspace{0.25em} \mathcal{F} ( K ) \longrightarrow
\tmmathbf{\tmop{Span}} \mathbf{}$, $\alpha : \hspace{0.25em} F \Rightarrow G$,
$\beta : \hspace{0.25em} H \Rightarrow G$ and $\mathcal{F} ( K ) \ni f : a
\rightarrow b$. We then have

\begin{center}
  $
\xymatrix @=5mm{
	Fa & 
        Ga \ar[l]_{\one} \ar[r]^{\alpha_a} 
        \xtwocell[1,1]{}\omit{^<6>{\alpha_f}} & 
        Ga &
        Ha \ar[l]_{\beta_a} \ar[r]^{\one}
        \xtwocell[1,-1]{}\omit{<-6>{\beta_f}} &
        Ha
	\\
        Ff \ar[u]^{(Ff)_1} \ar[d]_{(Ff)_2} & & 
	Gf \ar[u]^{(Gf)_1} \ar[d]_{(Gf)_2} & &
        Hf \ar[u]_{(Hf)_1} \ar[d]^{(Hf)_2}
	\\
        Fb & 
	Fb \ar[l]_{\one} \ar[r]^{\alpha_b} & 
	Gb &
        Hb \ar[l]_{\beta_b} \ar[r]^{\one} &
        Hb
}
$

\end{center}

{\noindent}at $f$. Let $\wp \left( s, t \right)$ be the pullback data of
morphisms $s$ and $t$ with common codomain in a category with pullbacks (i.e.
the triple consisting of the pullback object and the projections). There is
the series
\begin{enumerate}
  \item $\left( Q', q_1', q_2' \right) \equallim^{\tmop{def} .} \wp \left(
  \alpha_a, \left( Gf \right)_1 \right)$
  
  \item $\left( Q'', q_1'', q_2'' \right) \equallim^{\tmop{def} .} \wp \left(
  \beta_a, \left( Gf \right)_1 \right)$
  
  \item $\left( Q, q_1, q_2 \right) \equallim^{\tmop{def} .} \wp \left( q_2'
  \circ \alpha_f, q_2'' \circ \beta_f \right)$
  
  \item $\left( R, r_1, r_2 \right) \equallim^{\tmop{def} .} \wp \left(
  \alpha_a, \beta_a \right)$
  
  \item $\left( S, s_1, s_2 \right) \equallim^{\tmop{def} .} \wp \left(
  \alpha_b, \beta_b \right)$
\end{enumerate}
of pullback data in $\mathbb{B}$ and the pseudo-functor resulting from pulling
back $\alpha$ along $\beta$ evaluates at $f$ as

\begin{center}
  $\xymatrix@=5mm{
& Q 
\ar[dl]_{<(Ff)_1 \: \circ \: q_1 , (Hf)_1 \: \circ \: q_2>}
\ar[dr]^{<(Ff)_2 \: \circ \: q_1 , (Hf)_2 \: \circ \: q_2>}
&
\\
R&&S
}$
\end{center}

{\noindent}The legs of this span are given by universal property.

\section{\label{sec-cubes}Cubical Sets}

Similarly to simplicial sets which are presheaves over the simplicial category
$\Delta$, cubical sets are presheaves over a category of ``ideal cubes''
$\Box$. They were introduced by Serre in his thesis back in the early 1950's
{\cite{serre-phd}}. They turn out to be more convenient than simplicial sets
in some situations, since their product is geometrically simpler. Particularly
for topologists of the Bangor school, those devices have been standard tools of
the trade since some 25 years {\cite{alg-cub}} {\footnote{This paper only
marks the beginning of a long series.}}.

\subsection{Cubical Sets}

\begin{definition}
  \label{def:A-cubical-set}A cubical set $K$ is a sequence
  
  \begin{center}
    $\xymatrix{
\cdots \;\; K_n \ar@<2.5ex>[r]^{\partial^{+}_{i}} 
\ar@<-2.5ex>[r]_{\partial^{-}_{i}} &
K_{n-1} \ar[l]^>>>>{\epsilon_i}\;\; \cdots
}$

  \end{center}
  
  of sets and functions where $1 \leq i \leq n$, subject to the cubical
  identities
  \begin{enumerateroman}
    \item $\partial_i^{\omega} \circ \partial_j^{\omega'} = \partial_{j -
    1}^{\omega'} \circ \partial_i^{\omega}$ \text{for all} $i < j$ and $\omega, \omega'
    \in \left\{ -, + \right\}$
    
    \item $\epsilon_i \circ \epsilon_j = \epsilon_{j + 1} \circ \epsilon_i$
    \text{for all} $i \leq j$
    
    \item $\partial_i^{\omega} \circ \epsilon_j = \left\{ \begin{array}{llcc}
      \epsilon_{j - 1} \circ \partial_i^{\omega} & \text{for all}\;\; i < j
      \;\;\text{and}\;\; \omega \in \left\{ -, + \right\} &  & \\
      \epsilon_j \circ \partial_{i - 1}^{\omega} & \text{for all}\;\; \textrm{} i >
      j \textrm{\;\;\text{and}\;\;} \omega \in \left\{ -, + \right\} &  & \\
      id & \text{for all}\;\; i = j &  & 
    \end{array} \right.$
  \end{enumerateroman}
  
\end{definition}

The $\partial$'s are called (positive respectively negative) \textit{faces}
while the $\epsilon$'s are called \textit{degeneracies}. Again in analogy to
simplicial sets, elements of $K_n$ are called $n$-cubes. A cube in the image
of a degeneracy is called {\tmem{degenerate}}. 1-faces of an $n$-cube $y$,
obtained from the latter by applying a composite of $n - 1$ face maps, are
called $y$'s {\tmem{edges}}.

\begin{definition}
  \label{def-box}Let
  \begin{itemizeminus}
    \item $2 \equallim \{ 0, 1 \}$ and $\tmmathbf{1} = \{ \ast \}$;
    
    \item $\varepsilon \equallim^{\tmop{def} .} !_2 : 2 \rightarrow
    \tmmathbf{1}$;
    
    \item $\delta^{^-} : \tmmathbf{1} \rightarrow 2 \;\;\;\ast \mapsto 0$;
    
    \item $\delta^{^+_{}} : \tmmathbf{1} \rightarrow 2 \;\;\;\ast \mapsto 1$;
    
    \item $2^0 \equallim^{\tmop{def} .} \tmmathbf{1}$ and $2^p \boxdot 2^q
    \equallim^{\tmop{def} .} 2^{p + q}$.
  \end{itemizeminus}
  The category $\Box$ has $\{ 2^m | m \in \mathbb{N} \}$ as its set of
  objects and is generated by the maps
  \begin{itemizeminus}
    \item $\delta_i^{\omega} \equallim^{\tmop{def} .} 2^{i - 1} \boxdot
    \delta^{\omega} \boxdot 2^{n - i} : 2^{n - 1} \longrightarrow 2^n$ for all
    $1 \leqslant i \leqslant n$ and $\omega \in \{ -, + \}$;
    
    \item $\varepsilon_i \equallim^{\tmop{def} .} 2^{i - 1} \boxdot
    \varepsilon \boxdot 2^{n - i} : 2^n \longrightarrow 2^{n - 1}$ for all $1
    \leqslant i \leqslant n$
  \end{itemizeminus}
\end{definition}

The $\delta_i^{\omega}$ are called {\tmem{cofaces}} while the $\varepsilon_i$
are called {\tmem{codegeneracies}}. $\Box$ is strict monoidal (a PRO in fact),
yet the tensor $\boxdot$ is not a product although it is induced by the product
in $\tmmathbf{\tmop{Set}}$ {\cite{gr-mau}}. We obtain the {\tmem{cocubical
identities}}
\begin{enumerateroman}
  \item $\delta_j^{\omega'} \circ \delta_i^{\omega} = \delta_i^{\omega} \circ
  \delta_{j - 1}^{\omega'}$ for all $i < j$ and $\omega, \omega' \in \{ -, +
  \}$;
  
  \item $\varepsilon_j \circ \varepsilon_i = \varepsilon_i \circ
  \varepsilon_{j + 1}$ for all $i \leqslant j$;
  
  \item $\varepsilon_j \circ \delta_i^{\omega} = \left\{ \begin{array}{ll}
    \delta_i^{\omega} \circ \varepsilon_{j - 1} & i < j\\
    \delta^{\omega}_{i - 1} \circ \varepsilon_j & i > j\\
    \tmop{id} & i = j
  \end{array} \right.$ for all $\omega \in \{ -, + \}$
\end{enumerateroman}
from the two basic two ones $\varepsilon \circ \delta^{\omega} = \tmop{id}$
for $\omega \in \{ -, + \}$, so a cubical set $K$ is indeed a presheaf in
$\tmmathbf{\tmmathbf{\tmmathbf{\tmop{cSet}}}} \equallim^{\tmop{def} .}
\tmmathbf{\tmop{Set}}^{\Box^{\tmop{op}}}$. Morphisms in this category are
called {\tmem{cubical maps}}. As a matter of notation, $K_p$ stands for $K (
2^p )$ while $t_p$ stands for the component $t_{2^p}$ of the cubical map $t$
at $2^p$.  We write $\tmmathbf{\tmop{cSet}_{\bullet}}$ for the category of
pointed cubical sets.

\begin{lemma}
  \label{lem-canon} {\dueto{Grandis and Mauri {\cite{gr-mau}}}}Each composite
  of cofaces and codegeneracies has a canonical form
  \[ \delta_{j_1}^{\omega_1} \circ \cdots \circ \delta_{j_s}^{w_s} \circ
     \varepsilon_{i_1} \cdots \circ \varepsilon_{i_t} : 2^m \rightarrow 2^{m -
     t} \rightarrow 2^n \]
  where $1 \leqslant i_1 < \cdots < i_t$, $n \geqslant j_1 > \cdots > j_s
  \geqslant n$ and $m - t = n - s \geqslant 0$. This canonical form is unique.
\end{lemma}

\begin{theproof}
  The cocubical identities act as rewriting rules from left to right. 
\end{theproof}

\begin{remark}
  \label{rem-canon}Obviously, each composite of faces and degeneracies of a
  cubical set has a canonical form too. The factors just reverse.
\end{remark}

\subsection{Skeletons and Coskeletons}

Let $\Box_n \subset \Box$ be the full subcategory with objects $\{ 2^0 \cdots
2^n \}$, $J^{\tmop{op}} : \Box_n^{\tmop{op}} \hookrightarrow \Box^{\tmop{op}}$
be the dual of the inclusion and $\tmop{tr}_n \equallim^{\tmop{def} .} (_- )
\circ J^{\tmop{op}} : \tmmathbf{\tmop{Set}}^{\Box^{\tmop{op}}} \longrightarrow
\tmmathbf{\tmop{Set}^{\Box_n^{\tmop{op}}}}$ be the ($n$-){\tmem{truncation}}.
Since left and right Kan extensions to $\tmmathbf{\tmop{Set}}$ always exist,
there are adjunctions
\[ \tmop{sk}_n \equallim^{\tmop{def} .} \tmop{Lan}_{J^{\tmop{op}}} (_- )
   \dashv \tmop{tr}_n \dashv \tmop{cosk}_n \equallim^{\tmop{def} .}
   \tmop{Ran}_{J^{\tmop{op}}} (_- ) \]
In analogy to the simplicial situation, a cubical set in the image of
$\tmop{sk}_n$ is called {\tmem{n-skeletal}}. Given an arbitrary cubical set
$K$, its image under $\tmop{Sk}_n \equallim^{\tmop{def} .} \tmop{sk}_n \circ
\tmop{tr}_n$ is called its \tmem{n}th skeleton. It is easy to see that such
a cubical set has only degenerate cubes in dimensions above $n$.

On the other hand, a cubical set in the image of $\tmop{cosk}_n$ is called
{\tmem{n-coskeletal}} while applying $\tmop{Cosk}_n \equallim^{\tmop{def} .}
\tmop{cosk}_n \circ \tmop{tr}_n$ yields the {\tmem{n-coskeleton}}
{\footnote{Since this piece of terminology is chosen in analogy to the
simplicial case, the ``co'' is on the wrong side due to a ``historical
accident''  {\cite{duskin-simtriple}}.}}. Obviously, an $n$-coskeletal cubical
set is isomorphic to its $n$-coskeleton. 

The representable presheaf $\Box [ k ] \equallim^{\tmop{def} .} \Box (_-, 2^k
)$ is called the {\tmem{standard k-cube}}. $\Box [ k ]$ is $k$-skeletal with
precisely one non-degenerated cube in dimension $k$, namely $s_k
\equallim^{\tmop{def} .} \text{{\tmem{id}}}_{2^k}$. The faces of this cube are
$\partial_i^{\omega} ( s_k ) = \delta_i^{\omega} \in \Box ( 2^{k - 1}, 2^k )$
for $1 \leqslant i \leqslant k$ and $\omega \in \{ -, + \}$. The ($k-1$)th 
skeleton $\partial \Box [ k ]$ of the standard
$k$-cube is called {\tmem{boundary}}.

\begin{definition}
  \label{def-cubker}Suppose $Y \in
  \tmmathbf{\tmop{Set}}^{\Box_n^{\tmop{op}}}$. Its {\tmem{cubical kernel}}  is
  a set $Y^{\star}_{n + 1} \text{}^{}$ equipped with a family of maps
  \[ d_i^{\omega} : Y_{n + 1}^{\star} \rightarrow Y_n \]
  for all $1 \leqslant i \leqslant n + 1$ and $w \in \{ -, + \}$, such that
  the cubical identities hold and such that  $Y_{n + 1}^{\star}$ satisfies the
  following universal property. Given any set $Z$ with maps $z_i^{\omega} : Z
  \rightarrow Y_n$ such that the cubical identities hold, there is a unique
  map $z : Z \rightarrow Y_{n + 1}^{\star}$ such that
  \[ d_i^{\omega} \circ z = z_i^{\omega} \]
  for all $1 \leqslant i \leqslant n + 1$ and $w \in \{ -, + \}$.
\end{definition}

\begin{lemma}
  \label{lem-cubker}The cubical kernel exists for all $n \in \mathbb{N}$ and
  all  $Y \in \tmmathbf{\tmop{Set}}^{\Box_n^{\tmop{op}}}$.
\end{lemma}

\begin{theproof}
  $Y_{n + 1}^{\star} \equallim^{\tmop{def} .}  \left\{ ( y_i^{\omega} ) \in
  \prod_{1 \leqslant i \leqslant n + 1} \prod_{\omega \in \{ -, + \}} Y_n |
  \forall 1 \leqslant i < j \leqslant n + 1. \partial_i^{\omega} (
  y_j^{\omega'} ) = \partial_{j - 1}^{\omega'} ( y_i^{\omega} ) \right\} $ 
\end{theproof}

Elements of  $Y_{n + 1}^{\star}$ are called {\tmem{n-shells}}
{\cite{alg-cub}}. Any boundary is a shell.

\begin{proposition}
  \label{prop-cosk} The following are equivalent:
  \begin{enumerateroman}
    \item $X$ is n-coskeletal;
    
    \item any boundary in $X$ has a unique ``lifting'', i.e. any cubical map
    $\partial \Box [ k ] \longrightarrow X$ uniquely extends through the
    boundary inclusion:

    \begin{center}
      $\xymatrix{
\partial \Box [k] \ar[r] \ar@{^{(}->}[d] & X \\
\Box [k] \ar[ur]_{!} & \\
}$

    \end{center}
    
    {\noindent}for all $k > n$;
    
    \item $X_k$ is an iterated cubical kernel for all $k > n$.
  \end{enumerateroman}
\end{proposition}

\begin{theproof}
  $( i ) \Rightarrow ( ii )$ By hypothesis there is $X' :
  \Box_{_n}^{\tmop{op}} \rightarrow \tmmathbf{\tmop{Set}}$ such that $X \cong
  \tmop{cosk}_n ( X' ) .$ Let
  \[ \pi^{( m )} : 2^m \downarrow J \rightarrow \tmmathbf{\tmop{Set}} \]
  be the projection functor. We have $X_m = \lim ( X' \circ \pi^{( m )} )$.
  Unraveling yields
  \[ X_m = \left\{ ( x_s )_{} \in \prod_{\tmscript{\begin{array}{l}
       s : 2^p \rightarrow 2^m\\
       p \leqslant n
     \end{array}}} X' ( \tmop{dom}(s) ) \;|\;  \begin{array}{l}
       \forall p, q \leqslant n, h : 2^q \rightarrow 2^p, s : 2^p \rightarrow
       2^m, t : 2^q \rightarrow 2^m .\\
       t = s \circ h \Rightarrow X' ( h ) ( x_s ) = x_t
     \end{array} \right\}  \]
  {\noindent}Let $k > n$. A cubical map $\partial \Box [ k ] \longrightarrow
  X$ amounts to a choice of elements
  \[ ( Y_u^{1^-} ), ( Y_u^{1^+} ), \ldots, ( Y_u^{k^-} ), ( Y_u^{k^+} ) \in
     X_{k - 1} \]
  such that
  \[ Y_{\delta_i^{\omega} \circ f}^{j^{\omega'}} = Y_{\delta_j^{w'} \circ
     f}^{( i + 1 )^{\omega}} \]
  for all $f : 2^p \rightarrow 2^{k - 2}$, $j \leqslant i$ and $\omega,
  \omega' \in \{ -, + \}$. Let $p \leqslant n$ and $s : 2^p \rightarrow 2^k$.
  The latter has a canonical form by lemma \ref{lem-canon}. Since $k > n$,
  this canonical form has at least one factor which is a coface so $s =
  \delta_i^{\omega} \circ u$ for some $u : 2^p \rightarrow 2^{k - 1}$ and
  $\delta_i^{\omega} : 2^{k - 1} \rightarrow 2^k$. The desired unique boundary
  extension is then given by the element $( x_s ) \in X_k$ with coordinate at
  $s$
  \[ x_s \equallim^{} y_u^{i^{\omega}} \]
  $( ii ) \Rightarrow ( iii )$ Iterating
  the cubical kernel yields a cubical set. Given $k > n$, faces are given by
  the  projections, i.e. $\partial_i^{\omega} \equallim^{\tmop{def} .}
  \pi_{\omega} \circ \pi_i$. Degeneracies are given by
  \[ \epsilon_j ( Y )^{\omega}_i = \left\{ \begin{array}{ll}
       ( \epsilon_{j - 1} \circ \partial_i^{\omega} ) ( Y ) & i < j\\
       ( \epsilon_j \circ \partial_{i - 1}^{\omega} ) ( Y ) & i > j\\
       Y & i = j
     \end{array} \right. \]
  (c.f. {\cite{alg-cub}}). But an element of a cubical kernel amounts to a
  cubical map $\partial \Box [ k ] \longrightarrow X$.
  
  {\noindent}$( iii  ) \Rightarrow ( i )$ It
  is not hard to see that $X_k \cong \tmop{Cosk}_n ( X )_k$ enjoys the
  universal property of definition \ref{def-cubker} for any $k > n$.
\end{theproof}

The 1-truncation of a cubical set is a {\tmem{reflexive graph}}, i.e. a graph
with a distinguished loop at every vertex. The other way round, a reflexive
graph is a 1-skeletal cubical set. We use both terms interchangeably. As
customary, we do not make a notational difference between a set $\Sigma$ and a (here
reflexive) graph $G$ with one vertex and such that $G_1 = \Sigma$.

\subsection{Cubical Nerve and Categorification}
\label{sec-2-cubnerve}
\begin{lemma}
  {\dueto{Kan {\cite{kan1}}}} \label{lem-dex}Let $F : \mathbb{C} \rightarrow
  \mathbb{A}$ be a functor, $y : \mathbb{C} \longrightarrow
  \widehat{\mathbb{C}}$ be the Yoneda embedding and
  \[ \begin{array}{llll}
       F_{\ast} : & \mathbb{A} & \longrightarrow & \widehat{\mathbb{C}}\\
       & A & \longmapsto & \mathbb{A} ( F_-, A )
     \end{array} \]
  If it exists, $\text{Lan}_y F$ is left adjoint to $F_{\ast}$.
\end{lemma}

\begin{proposition}
  \label{prop-cubnerve}Let $J = - \longrightarrowlim^t +$ be the interval
  category, $d_{i, p}^{\omega} : J^{p - 1} \rightarrow J^p$ the functor
  inserting $\omega \in \{ -, + \}$ respectively $\tmop{id}_{\omega}$ at $i$th
  coordinate and $e_{i, p} : J^p \longrightarrow J^{p - 1}$ the functor
  deleting the $i$th coordinate. These data organize to a functor
  \[ \begin{array}{llll}
       c : & \Box & \longrightarrow & \tmmathbf{\tmop{Cat}}\\
       & 2^p & \longmapsto & J^p\\
       & \delta_{i, p}^{\omega} & \longmapsto & d_{i, p}^{\omega}\\
       & \varepsilon_{i, p} & \longmapsto & e_{i, p}
     \end{array} \]
  Let $y : \Box \longrightarrow \tmmathbf{\tmop{Set}^{\Box^{\tmop{op}}}}$ be
  the Yoneda embedding. There is the adjunction
  \[ \tmop{Lan}_y c \dashv c_{\ast} \]
\end{proposition}

\begin{theproof}
  By lemma \ref{lem-dex}.
\end{theproof}

\begin{definition}
  With the notation of proposition \ref{prop-cubnerve}, $\mathcal{C}
  \equallim^{\tmop{def} .} \tmop{Lan}_y c$ is called {\tmem{categorification
  while}} $\mathcal{N} \equallim^{\tmop{def} .} c_{\ast}$ is called
  {\tmem{cubical nerve}}.
\end{definition}

\begin{lemma}
  \label{lem-cubpaths}Let K be a cubical set and $y \in K_n$. Let $f, g \in
  K_1$ be $y$'s edges with canonical forms $f = ( \partial_{i_{n -
  1}}^{\omega_{n - 1}} \circ \cdots \circ \partial_{i_1}^{\omega_1} ) ( y )$
  and $g = ( \partial_{j_{n - 1}}^{\upsilon_{n - 1}} \circ \cdots \circ
  \partial_{j_1}^{\upsilon_1} ) ( y )$. The following are equivalent:
  \begin{enumerateroman}
    \item $\partial^- ( g ) = \partial^+ ( f )^{}$  ;
    
    \item there is precisely one $1 \leqslant k \leqslant n - 1$ such that
    $\forall 1 \leqslant l \leqslant n - 1. i_k \neq j_l$ and $\omega^k = -$ 
    .
  \end{enumerateroman}
\end{lemma}

\begin{corollary}
  There are exactly $n!$ paths from $\tmop{dom}_n ( y )$ to $\tmop{cod}_n ( y
  )$. 
\end{corollary}

\begin{proposition}
  \label{prop-categorif}Under the notation of lemma \ref{lem-cubpaths}, let
  \[ \tmop{dom}_n \left( y \right) \deq ( \underbrace{\partial_0^- \circ
     \cdots \circ \partial_0^-}_{n \times} ) ( y ) \]
  and
  \[ \tmop{cod}_n \left( y \right) \deq ( \partial_0^+ \circ \cdots \circ
     \partial_{n - 1}^+ ) ( y ) \]
  Suppose $P_y$ be the set of paths from $\tmop{dom}_n ( y )$ to $\tmop{cod}_n
  ( y )$. The categorification $\mathcal{C}( K )$ of a cubical set $K$ has the
  set of objects $\mathcal{C}( K )_0 = K_0$. It is generated by the reflexive
  graph $\tmop{tr}_1 ( K )$ and is subject to the {\tmem{cubical relations}}
  \[ \forall r, s \in P_y . r = s \]
  for all non-degenerate $y \in K_n$ in dimensions $n > 1$. The action of
  $\mathcal{C}$ on a cubical map $t$ is determined by $t_1$.
\end{proposition}

Proposition \ref{prop-categorif} is essentially a consequence of lemma
\ref{lem-cubpaths}. The details of the proofs consist of tedious, yet entirely
standard combinatorics.

\begin{remark}
  \label{rem-length}The $n$! paths ``around a cube'' are all different, yet 
their
  categorifications do not in general have the same length. Consider for
  instance the non-degenerate cube

  \begin{center}
    $
\xymatrix {b \ar[r]^{\epsilon (b)} \ar@{}[dr]|-{K} & b
\\
a \ar[u]^{h} \ar[r]_f & c \ar[u]_g
}
$

  \end{center}

  {\noindent}In particular, given a cubical set $K$ such that $K_0 =
  1_{\tmmathbf{\tmop{Set}}}$, we have
  \[ (\mathcal{C} \circ \tmop{Cosk}_1 ) ( K ) \cong 1_{\tmmathbf{\tmop{Cat}}}
  \]
  since in this case {\tmem{all}} paths are equated.
\end{remark}

\begin{definition}
  \label{def-cubobj}A cocubical object in category $\mathbb{C}$ is a functor
  $\text{$\Box \longrightarrow \mathbb{C}$}$. A cubical object in a category
  $\mathbb{C}$ is a functor $\Box^{\tmop{op}} \longrightarrow \mathbb{C}$.
\end{definition}

\begin{proposition}
  Let $\mathbb{C}$ be locally small. $C \in \mathbb{C}$ and a cocubical
  object $K : \Box \longrightarrow \mathbb{C}$ determine a cubical set
  \[ \tmop{Cubes}_K ( C ) \equallim^{\tmop{def} .}  \mathbb{C} ( K (_- ), C )
  \]
\end{proposition}

\begin{theproof}
  A cubical set is just a presheaf on $\Box$.
\end{theproof}

\begin{proposition}
  \label{prop-rgrcubes}Let $C = - \longrightarrowlim^c +$ be the interval
  reflexive graph and
  \[ K : ( \Box, \boxdot ) \longrightarrow ( \tmmathbf{\tmop{rGraph}}, \times
     ) \]
  be the monoidal functor such that
  \begin{itemizeminus}
    \item $K ( 2 ) = C$  ;
    
    \item $K ( \delta^- ) ( C ) = -$  ;
    
    \item $K ( \delta^+ ) ( C ) = +$  .
  \end{itemizeminus}
  Then
  \begin{enumerateroman}
    \item $\tmop{Cubes}_K ( R )_m = \tmmathbf{\tmop{rGraph}} ( C^m, R )$  ;
    
    \item there is the isomorphism of cubical sets
    \[ \tmop{cosk}_1 ( R ) \cong \tmop{Cubes}_K ( R ) \]
    for all $R \in \tmmathbf{\tmop{rGraph}}$.
  \end{enumerateroman}
\end{proposition}

\begin{theproof}
  
  \begin{enumerateroman}
    \item $\tmop{Cubes}_K ( R )_m = \tmmathbf{\tmop{rGraph}} ( C^m, R )$ since
    $K ( 2^m ) = C^m$  ;
    
    \item A morphism $h : C^2 \longrightarrow R$ determines the 2-shell
    \[ \text{$( ( h ( - c ), h ( c - ) ), ( h ( + c), h(c +) ) ) 
\in Y^{\star}_2 (
       R )$} \]
    Conversely, the 2-shell
    \[ ( ( y_1^-, y_2^- ), ( y_1^{+_{}}, y_2^+ ) ) \in Y^{\star}_2 ( R ) \]
    determines the morphism
    \[ \begin{array}{llll}
         l : & C^2 & \longrightarrow & R\\
         & - c & \longmapsto & y_1^-\\
         & c - & \longmapsto & y_2^-\\
         & + c & \longmapsto & y_1^+\\
         & c + & \longmapsto & y_2^+
       \end{array} \]
    The assertion follows by induction on $m$.
  \end{enumerateroman}
\end{theproof}

\begin{corollary}
  \label{cor-rgrcubes}Let
  \[ \tmop{NCubes}_K ( R ) \equallim^{\tmop{def} .}  \bigcup_{m \in
     \mathbb{N}} \{ h \in \tmop{Cubes}_K ( R )_m | h \;\; \tmop{non-degenerate} \} \]
  and $\sim^K$ be the congruence on $\mathcal{F} ( R )$ generated by
  \[ s \sim t \Longleftrightarrowlim^{\tmop{def} .} \exists h \in
     \tmop{NCubes}_K ( R ) . s \in \tmop{im} (\mathcal{F}( h) )
 \wedge t \in \tmop{im} ( \mathcal{F}(h)
     ) . \]
  Then
  \[ ( \mathcal{C} \circ \tmop{cosk}_1 ( R ) ) \cong \mathcal{F} ( R ) /
     \sim^K \]
\end{corollary}

\section{\label{sec-hdasync}Synchronization of Higher-Dimensional Automata}

For the computer scientist, cubical sets are a convenient tool to organize
independence relations among transitions {\cite{hha}}. Indeed, traditional
{\tmem{labeled transition systems}} over an alphabet $\Sigma$ only capture
the interleavings of potentially parallel computational paths. If one is
interested in ``true parallelism'', additional information needs to be
provided. A possible way to realize this program is to label transitions with
sequences over $\Sigma$.

\subsection{Automata with concurrency}

\begin{proposition}
  {\dueto{Goubault {\cite{eric:cmcim}}}} \label{def:bang-sigma}Given a set
  $\Sigma$ and $\alpha = \left( \alpha_1, \ldots, \alpha_n \right) \in
  \Sigma^{\ast}$, let $\left| \alpha \right| \equallim^{\tmop{def} .} n$ be
  $\alpha$'s length. Suppose from now on $\Sigma$ totally ordered and let
  \[ \alpha \models \left( \Sigma, \leq \right) \hspace{0.5em}
     \Longleftrightarrowlim^{\tmop{def} .} \hspace{0.5em} \alpha_1 \leq \cdots
     \leq \alpha_n \]
  Let further $\idle \nin \Sigma$, 
$\alpha = \left( \alpha_1, \ldots, \alpha_n\right) \in 
\left( \Sigma \cup \{ \idle \} \right)^{\ast}$,
  $\underline{} \left\| \alpha \right\| \subseteq \alpha$ be the word obtained
  from $\alpha$ by removing all the occurrences of $\star$ and
  \[ ! \Sigma_n \deq \left\{ \alpha \in \left( \Sigma \cup \left\{ \idle
     \right\} \right)^{\ast} \hspace{0.25em} \mid \hspace{0.25em} \left|
     \alpha \right| = n \hspace{0.5em} \wedge \hspace{0.5em} \left\| \alpha
     \right\| \models \left( \Sigma, \leq \right) \right\} \]
  $( ! \Sigma )_{n \in \mathbbm{N}}$ along with faces
  \[ \partial^-_i \left( \alpha_1, \ldots, \alpha_n \right) = \partial^+_i
     \left( \alpha_1, \ldots, \alpha_n \right) \equallim^{\tmop{def} .}
( \alpha_1, \ldots, \alpha_{i - 1}, \alpha_{i + 1}, \ldots, \alpha_n ) \]
  and degeneracies
  \[ \epsilon_i \left( \alpha_1, \ldots, \alpha_n \right) 
\equallim^{\tmop{def} .}
(\alpha_1, \ldots, \alpha_{i - 1}, \idle, \alpha_i, \ldots, \alpha_n) \]
  determine a cubical set.  
\end{proposition}

\begin{definition}
  A higher-dimensional automaton or HDA (over the alphabet $\Sigma$) is an
  object in the slice category $\tmmathbf{\tmop{cSet}} / ! \Sigma$. As a
  convention, let $! \Sigma_0 \equallim^{\tmop{def} .} \{ \star \}$. An HDA $t
  : T \longrightarrow ! \Sigma$ is {\tmem{n-skeletal}} provided $T$ is.
\end{definition}

{\noindent}Cubes in higher dimensions encode independence relations for each
arity while face maps encode the {\tmem{coherence}} of these relations.
Indeed, computational intuition says that if a set of transitions is pairwise
independent so each subset thereof has to be pairwise independent as well. Put
differently, an $n$-cube $k$ can be seen as an $n$-dimensional
transition from state $\tmop{dom}_n ( k )$ to state $\tmop{cod}_n ( k )$
(c.f. lemma \ref{lem-cubpaths}), the dimension being the number of processes
operating without interaction. A traditional labeled transition system is
thus a pointed $1$-skeletal HDA $t : T \longrightarrow ! \Sigma$ such that
$t_1$ is locally injective.

\subsection{Coskeletal HDA's}

Is there a notion of $n$-coskeleton suitable for HDA's, i.e. compatible with
labeling? Given an HDA  $t : T \longrightarrow ! \Sigma$, the first reaction
would off course be to consider $\tmop{Cosk}_n ( t )$. However, $! \Sigma$ is
obviously not $n$-coskeletal. Consider on the other hand the transition system

\begin{center}
  $
\xymatrix {b \ar[r]^{\alpha} & d \\
a \ar[u]^{\beta} \ar[r]_{\psi} 
& c \ar[u]_{\theta} 
}
$

\end{center}

{\noindent}It's underlying reflexive graph is a 1-shell, yet there is no
canonical way to label this 1-shell.

\begin{definition}
  \label{def-hdashell}Let $t : T \longrightarrow ! \Sigma$ be an HDA. A
  $\Sigma_n$-shell of $t$ is an element
  \[ y = ( ( y_1^-, \ldots, y_{n + 1}^- ), ( y_1^+, \ldots, y_{n + 1}^+ ) )
     \in T^{\star}_{n + 1} \]
  such that $t_n ( y_i^- ) = t_n ( y_i^+ )$ for all $1 \leqslant i \leqslant n
  + 1$. $T_{n + 1}^{\star \star}$ is the set of all $\Sigma_n$-shells and
  \[ \tmop{Cosk}_{\Sigma, n} ( T ) \rightarrowtail \tmop{Cosk}_n ( T ) \]
  is the subobject determined by the $T^{\star \star}_{n + j +
  1}$'s for all $j \in \mathbbm{N}$. This subobject is called the $n$-th
  {\tmem{$\Sigma$}}-coskeleton of $T$.
\end{definition}

In short, a $\Sigma$-coskeleton verifies the Kan condition of proposition
\ref{prop-cosk} for the ``good'' shells.

\begin{lemma}
  \label{lem-hdashell}Let $y$ be a $\Sigma_n$-shell of the HDA $t : T
  \longrightarrow ! \Sigma$. The sequence of labels
  \[ \left( t_n ( y_i^- ) \right) \in \prod_{1 \leqslant i \leqslant n + 1} !
     \Sigma_n \]
  is determined by $t_n ( y_1^- )$ and $( \partial_n^- \circ \cdots \circ
  \partial_{2^{}}^- \circ t_n ) ( y_2^- )$.
\end{lemma}

\begin{theproof}
  Let $t_n ( y_i^- ) = \alpha_i \equallim^{\tmop{def} .} ( \alpha_i^{{\small (
  1 )}}, \ldots, \alpha^{{\small ( n )}}_i )^{}$ for $1 \leqslant i \leqslant
  n + 1$. Since $t$ is a cubical map, we have
  \[ \partial_k^- ( \alpha_1 ) = \partial_1^- ( \alpha_{k + 1} ) \]
  for all $1 \leqslant k \leqslant n$, i.e.
  \[ ( \alpha_i^{{\small ( 2 )}}, \ldots, \alpha_i^{{\small ( n )}} ) = (
     \alpha_1^{{\small ( 1 )}}, \ldots, \alpha_1^{{\small ( k - 1 )}},
     \alpha_1^{{\small ( k + 1 )}}, \ldots, \alpha_1^{{\small ( n )}} ) \]
  for all $2 \leqslant i \leqslant n + 1$. Similarly,
  \[ \partial_i^- ( \alpha_i ) = ( \alpha_i^{{\small ( 1 )}}, \ldots,
     \alpha_i^{{\small ( k - 1 )}}, \alpha_i^{{\small ( k + 1 )}}, \ldots,
     \alpha_i^{{\small ( n )}} ) = ( \alpha_{i + 1}^{{\small ( 1 )}}, \ldots,
     \alpha_{i + 1}^{{\small ( k - 1 )}}, \alpha_{i + 1}^{{\small ( k + 1 )}},
     \ldots, \alpha_{i + 1}^{{\small ( n )}} ) = \partial_i^- ( \alpha_{i + 1}
     ) \]
  so in particular
  \[ \alpha_i^{{\small ( 1 )}} = \alpha_{i + 1}^{{\small ( 1 )}} \]
  for all $2 \leqslant i \leqslant n$.
\end{theproof}

\begin{proposition}
  \label{prop-hdashell}Let $t : T \longrightarrow ! \Sigma$ be an HDA. There
  is the HDA 
\[\tmop{Cosk}_{\Sigma, n} ( t ) : \tmop{Cosk}_{\Sigma, n} ( T )
  \longrightarrowlim ! \Sigma\]
given in dimension $k > n$ by
  \[ ( \tmop{Cosk}_{\Sigma, n} ( t ) )_k ( y ) \equallim^{\tmop{def} .} ( t_{k
     - 1} ( y_1^- ), ( \partial_n^- \circ \cdots \circ \partial_{2^{}}^- \circ
     t_{k - 1} ) ( y_2^- ) ) \]
\end{proposition}

\begin{theproof}
  By lemma \ref{lem-hdashell}.
\end{theproof}

When $n = 1$, we write $\tmop{Cosk}_{\Sigma} ( t ) : \tmop{Cosk}_{\Sigma} ( T
) \rightarrowlim ! \Sigma$ respectively $\tmop{cosk}_{\Sigma} ( r ) :
\tmop{cosk}_{\Sigma} ( R ) \rightarrow ! \Sigma$. Clearly, both lemma
\ref{lem-hdashell} and proposition \ref{prop-hdashell} could also be
formulated with the ``positive'' parts of the $\Sigma$-shells.

\subsection{Coskeletal Synchronization}

Synchronization of transition systems can be presented by a table also
(glamorously) called synchronization algebra {\cite{NielsenM:modc1}}. The
idea is to filter out and to relabel transitions out of a product. This
amounts to the construction of a pullback. Similarly, synchronization of HDA's
can be presented along those lines. We introduce here the simplest case called
{\tmem{1-coskeletal synchronization}}, the general treatment will appear
elsewhere. 


As the name suggests, 1-coskeletal synchronization means that everything is
determined by dimension 1. Suppose for instance $\Sigma = \{ \alpha, \beta,
\bar{\alpha}, \bar{\beta}, \tau \}$ and consider the synchronization table
\[ \begin{array}{|c||c|c|c|c|c|c|}
     \hline
     & \star & \alpha & \beta & \bar{\alpha} & \bar{\beta} & \tau\\
     \hline\hline
     \star & \star & \alpha & \beta & \bar{\alpha} & \bar{\beta} & \tau\\
     \hline
     \alpha & \alpha & \top & \top & \top & \top & \top\\
     \hline
     \beta & \beta & \top & \top & \top & \tau & \top\\
     \hline
     \bar{\alpha} & \bar{\alpha} & \tau & \top & \top & \top & \top\\
     \hline
     \bar{\beta} & \bar{\beta} & \top & \tau & \top & \top & \top\\
     \hline
     \tau & \tau & \top & \top & \top & \top & \top\\
     \hline
   \end{array}  \]
indexed by $\Sigma \cup \{ \star \}$ and with entries in $\Sigma \cup \{
\star, \top \}$. This table prescribes the following ``synchronized product''
of HDA's:
\begin{itemizeminus}
  \item an idle transition (with label $\star$) is synchronized with
  any transition and the result is relabeled as the latter;
  
  \item $\alpha$-transitions are synchronized with $\bar{\alpha}$-transitions
  and the result is a $\tau$-transition; similarly for labels $\beta$ and
  $\bar{\beta}$;
  
  \item all other pairs of transitions are not synchronized (i.e. filtered
  out); this is indicated by $\top$ and suggests that this information shall
  propagate to higher dimensions.
\end{itemizeminus}
Consider the (2-skeletal) HDA
\[ \begin{array}{llll}
     u : & \Box [ 2 ] & \longrightarrow & ! \Sigma\\
     & s_2 & \longmapsto & ( \alpha, \beta )
   \end{array} \]
depicted as

\begin{center}
  $
\xymatrix {b \ar[r]^{\alpha} & d \\
a \ar[u]^{\beta} \ar[r]_{\alpha} \ar@{}[ur]|{(\alpha,\beta)}
& c \ar[u]_{\beta} 
}
$

\end{center}

{\noindent}The picture does not include the degenerate cubes. It illustrates
the fact that labeling the standard 2-cube with $( \alpha, \beta )$ implies
the labels of its faces, this since $u$ is technically a cubical map. Consider
further the (1-skeletal) HDA
\[ \begin{array}{llll}
     v : & \Box [ 1 ] & \longrightarrow & ! \Sigma\\
     & s_1 & \longmapsto & \bar{\beta}
   \end{array} \]
depicted as
\[ x \longrightarrowlim^{\bar{\beta}} y \]
The synchronized product $u \boxast v$ of $u$ and $v$ with respect to the
above table is

\begin{center}
  $
\xymatrix {b \ar[r]^{\alpha} & d \\
a \ar[u]^{\tau} \ar[r]_{\alpha} \ar@{}[ur]|{(\alpha,\tau)}
& c \ar[u]_{\tau} 
}
$

\end{center}

\begin{lemma}
  \label{lem-topos}Let $i : S \rightarrowtail A \times B$ be a relation in a
  topos. The pushout square determined by the projections $\pi_1 \circ i$ and
  $\pi_2 \circ i$ is a pullback square.
\end{lemma}

\begin{theproof}
  Let $s : A \times B \rightarrow \Omega$ be the classifying predicate of $S$
  and $P \equallim^{\tmop{def} .} A +_{{\small S}} B$ with coprojections $j_1$
  and $j_2$. Then
  \[ A \times_{{\small P}} B \cong \{ ( a, b ) \in A \times B | j_1 ( a ) =
     j_2 ( b ) \} = \{ ( a, b ) \in A \times B | s ( a, b ) \} \cong S . \]
\end{theproof}

\begin{proposition}
  \label{prop-pushpull}Let $s : S \rightarrowlim ! \Sigma$ be an HDA and
  $\check{s} : \tmop{tr}_1 ( S ) \rightarrow \Sigma_s$ be the morphism of
  reflexive graphs given by the image factorization
  
  \begin{center}
    $
\xymatrix {tr_1(S)  
 \ar@{->>}[dr]^{\check{s}} \ar[dd]_{tr_1(s)} & \\
&\Sigma_s \ar@{>->}[dl] \\
tr_1(!\Sigma) 
}
$

  \end{center}
  
  {\noindent}in the topos $\tmmathbf{\tmop{rGraph}}$. Let $\tmmathbf{s} (_-,_-
  ) : \left( \Sigma \cup \{ \star \} \right)^2 \rightarrow \Sigma \cup \{
  \star, \top \}$ be a synchronization table. Given a further HDA $t : T
  \rightarrowlim ! \Sigma$, let
  \[ \Upsilon'_{s, t} \equallim^{\tmop{def} .}  \left\{ ( \theta, \psi ) \in
     \Sigma_s \times \Sigma_t \;|\;
 \tmmathbf{s} ( \theta, \psi ) \neq \top \wedge
     \theta \neq \star \wedge \psi \neq \star \right\} \]
  and
  \[ \Upsilon_{s, t} \equallim^{\tmop{def} .} \Upsilon'_{s, t} \cup \left( (
     \{ \star \} \times \Sigma_t ) \backslash ( \{ \star \} \times \pi_2 (
     \Upsilon'_{s, t} ) ) \right) \cup ( ( \Sigma_s \times \{ \star \} )
     \backslash ( \pi_1 ( \Upsilon'_{s, t} ) \times \{ \star \} ) ) ) \]
  Under these assumptions, the pushout square
  
  \begin{center}
    $
\xy
\xymatrix {\Upsilon_{s,t}
\ar[r]^{\pi_2} \ar[d]_{\pi_1}
&
\Sigma_t \ar[d]
\\
\Sigma_s \ar[r] & \Sigma_{s,t}
}
\POS(11.5,-9);\POS(11.5,-11.5) \connect@{-}
\POS(11.5,-9);\POS(14,-9) \connect@{-}
\endxy
$

  \end{center}
  
  {\noindent}is also a pullback square.
\end{proposition}

\begin{theproof}
  By lemma \ref{lem-topos}.
\end{theproof}

If $s$ is 1-skeletal, we slightly abuse notation and write $s : S \rightarrow
\Sigma_s$ for $\check{s}$ of proposition \ref{prop-pushpull}.

\begin{definition}
  \label{def-syncprod}Under the notation of proposition of proposition
  \ref{prop-pushpull}, let
  \[ S \boxtimes_{\tmmathbf{s}} T \; \equallim^{\tmop{def} .}\;
 \tmop{tr}_1 ( S )
     \times_{\Sigma_{s, t}} \tmop{tr}_1 ( T ) \]
  and $\langle s, t \rangle$ be the comparison morphism in

  \begin{center}
    $
\xy
\xymatrix {S  \boxtimes_{\mathbf{s}}  T 
\ar[rr] \ar[dd] \ar@{.>}[dr]^{\langle s,t \rangle}
&& \mathrm{tr}_1(T) \ar[d]^{\check{t}}
\\
&\Upsilon_{s,t}
\ar[r]^{\pi_2} \ar[d]_{\pi_1}
&
\Sigma_t \ar[d]^{j_{s,t}^T}
\\
\mathrm{tr}_1(S) \ar[r]_{\check{s}} & \Sigma_s
 \ar[r]_{j_{s,t}^S} & \Sigma_{s,t}
}

\POS(6.0,-4.9);\POS(6.0,-2.4) \connect@{-}
\POS(6.0,-4.9);\POS(3.5,-4.9) \connect@{-}

\POS(24,-19);\POS(24,-16.5) \connect@{-}
\POS(24,-19);\POS(21.5,-19) \connect@{-}

\endxy
$

  \end{center}
  
  {\noindent}The formula for the synchronized product is
  \[ s \boxast_{\tmmathbf{s}} t \;\equallim^{\tmop{def} .}\;
 \tmop{cosk}_{\Sigma}
     \left( \tmmathbf{s} (_-,_- ) |_{\Upsilon_{s, t}} \circ \langle s, t
     \rangle \right)  \]
\end{definition}

\subsection{Categorification of Synchronized HDA'a}

\begin{definition}
  \label{def-graphcube}Let $C = - \longrightarrowlim^{\kappa} +$ be the
  interval reflexive graph and $m \in \mathbbm{N}$. An $m$-cube in a reflexive
  graph $K$ is a morphism $c : C^m \rightarrow K$ (c.f. proposition
  \ref{prop-rgrcubes}). An $m$-cube is {\tmem{rigid}} if  $e \neq \tmop{id}
  \Rightarrow c ( e ) \neq \tmop{id}$ for all edges $e \in ( C^m )_1$ and
  {\tmem{contractible}} if $c ( e ) = \tmop{id}$ for all edges $e \in ( C^m
  )_1$. K is {\tmem{acubic}} if all the $m$-cubes in $K$ are contractible for
  all $m \in \mathbbm{N}$. 
\end{definition}

In definition \ref{def-graphcube}, $\tmop{id}$ stands off course for a
distinguished self-loop in a reflexive graph. We stick to this abuse of
terminology and notation. We write $\partial^- ( u )$ for $\tmop{dom} ( u )$
respectively $\partial^+ ( u )$ for $\tmop{cod} ( u )$ when convenient and use
the ``turtle graphics'' notation for vertices and edges of $C^m$, e.g.

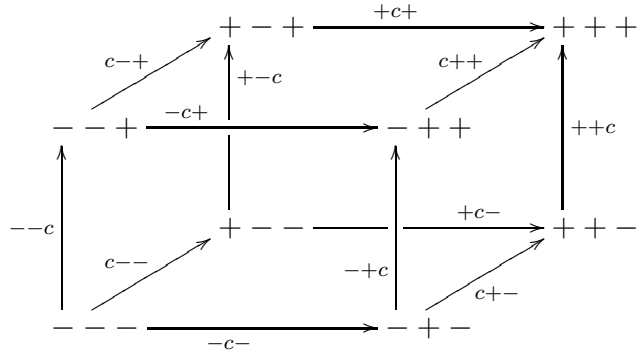
\begin{center}
  $
\xymatrix @*[r] {& +-+ \ar[rr]^{+c+} & & +++ 
\\
--+ \ar[ru]^{c-+} \ar[rr]^<<<<<<{-c+} && -++ \ar[ru]^{c++} & 
\\
& +-- \ar'[r][rr]^{+c-} \ar'[u][uu]_{+-c} && ++- \ar[uu]_{++c}
\\
--- \ar[uu]^{--c} \ar[rr]_{-c-} \ar[ru]^{c--} && -+-\ar[ru]_{c+-} 
\ar[uu]^<<<<<<{-+c} &
}
$

\end{center}

\begin{theorem}
  \label{theo-catsynchro}Let $\Sigma$ be an alphabet, $\mathcal{F :
  \tmmathbf{\tmop{rGraph}} \rightarrow \tmmathbf{\tmop{Cat}}}$ be the ``free''
  functor and $t_i : T_i \rightarrowlim \Sigma$ be morphisms of reflexive
  graphs for $1 \leqslant i \leqslant m$. Suppose that $T_i$ is acubic for
  all $1 \leqslant i \leqslant m$. Let $\Sigma_{i, i + 1}
  \equallim^{\tmop{def} .} \Sigma_{t_i, t_{i + 1}}$ for all $1 \leqslant i
  \leqslant m - 1$. Then
  \[ ( \mathcal{C} \circ \tmop{cosk}_{\Sigma} ) ( T_1 \boxtimes \cdots
     \boxtimes T_m ) \cong 
\mathcal{F} ( T_1 ) 
\times_{\mathcal{F} (\Sigma_{1, 2} )}
  \mathcal{F} ( T_2 ) \times_{ \mathcal{F}
     ( \Sigma_{2, 3} )} \cdots 
\times_{\mathcal{F} ( \Sigma_{m - 1,m} )} 
\mathcal{F} ( T_m ) \]
\end{theorem}

\begin{theproof}
  Let
  \[ t_i^- \equallim^{\tmop{def} .} \left\{\begin{array}{ll}
         j^{T_i}_{t_{i - 1}, t_i} \circ t_i
        & 1 < i \leqslant m \\ \\
       j^{T_1}_{t_1, t_2} \circ t_1 & i = 1
     \end{array}\right. \]
  and
  \[ t_i^+ \equallim^{\tmop{def} .} \left\{
       \begin{array}{ll}
         j^{T_i}_{t_i, t_{i + 1}} \circ t_i & 1 \leqslant i < m \\ \\
         j^{T_m}_{t_{m - 1}, t_m} \circ t_m & i = m
     \end{array}\right. \]
  $T_1 \boxtimes \cdots \boxtimes T_m$ is the limit of the diagram

  \begin{center}
    $
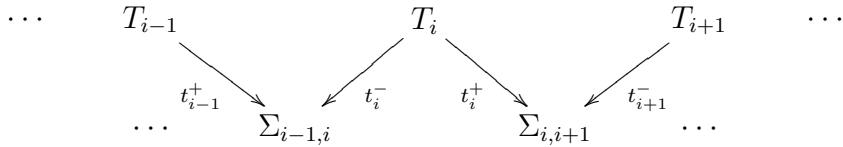
\xymatrix { \cdots & T_{i-1} \ar[dr]_{t_{i-1}^+} && 
T_i \ar[dl]^{t_i^-} \ar[dr]_{t_i^+}
&&
T_{i+1} \ar[dl]^{t_{i+1}^-} & \cdots
\\
& \cdots &
\Sigma_{i-1,i} &&
\Sigma_{i,i+1} &
\cdots &
}
$

  \end{center}
  
  {\noindent}hence
\[ 
\begin{array}{ccl}
( e_1, \ldots, e_m ) \in T_1 \boxtimes \cdots \boxtimes T_m &
     \Longleftrightarrowlim &
( e_1, \ldots, e_m ) \in T_1 \times \cdots \times 
     T_m  \;\;\;\;\;(\ast ) \\
&&\;\; \wedge \\
&&\forall 1 \leqslant i < m. t^+ ( e_i ) = t^- ( e_{i + 1} ) 
\end{array} 
\]
  Suppose $\text{$p \leqslant m$}$ and let $\{ i_1, \ldots, i_p \} \subseteq
  \{ 1, \ldots, m \}$ be a set of indices. Let $( u_{i_q} ) \in \prod_{1
  \leqslant q \leqslant p} ( T_{i_q} )_{_{_1}}$ be a family of edges such that
  $u_{i_q} \neq \tmop{id}$ and
  \[ t_{i_q}^- ( u_{i_q} ) = t_{i_q}^+ ( u_{i_q} ) = \tmop{id} \]
  for all $1 \leqslant q \leqslant p$. Let $( x_j ) \in \prod_{j \nin \{ i_1,
  \ldots, i_p \}} T_j$ be a family of vertices. Analyzing the above wide
  pullback, one sees that the data $( u_{i_q} )$ and $( x_j )$ determine a
  rigid $p$-cube
  \[ c : \text{$C^p \rightarrow T_1 \boxtimes \cdots \boxtimes T_m$} \]
  such that $c ( \omega_1 \cdots \omega_{r - 1} \kappa \omega_{r + 1} \cdots
  \omega_p ) = ( \tmop{id} ( y_1 ), \ldots, \tmop{id} ( y_{i_r - 1} ),
  u_{i_r}, \tmop{id} ( y_{i_r + 1} ), \ldots, \tmop{id} ( y_m ) )$ with
  \[ y_j \equallim^{\tmop{def} .}  \left\{\begin{array}{l}
       \begin{array}{ll}
         _{} \partial^{\omega_q} ( u_{i_q} ) & \exists q \in \{ 1, \ldots, r -
         1, r + 1, \ldots, p \} . \;j = i_q\\
         x_j & 
       \end{array}
     \end{array}\right. \]
  Conversely, any non-contractible cube in $T_1 \boxtimes \cdots \boxtimes
  T_m$  arises as above since the $T_i$'s are acubic by hypothesis. In
  particular, any $p$-cube in $T_1 \boxtimes \cdots
  \boxtimes T_m$ is contractible for $p > m$ and
  \[ \tmop{cosk}_{\Sigma} ( T_1 \boxtimes \cdots \boxtimes T_m ) \cong
     \tmop{cosk} ( T_1 \boxtimes \cdots \boxtimes T_m ) \;\;\;( \ast \ast ) \]
  Suppose now $( e_1, \ldots, e_m ) \in T_1 \boxtimes \cdots \boxtimes T_m$
  and let $( e_i ) \in \mathcal{F} ( T_i )$ be the path of length $1$ determined
  by $e_i$. We have
  \[ ( ( e_1 ), \ldots, ( e_m ) ) \in 
( 
\mathcal{F} ( T_1 ) 
\times_{\mathcal{F} (\Sigma_{1, 2} )}
  \mathcal{F} ( T_2 ) \times_{ \mathcal{F}
     ( \Sigma_{2, 3} )} \cdots 
\times_{\mathcal{F} ( \Sigma_{m - 1,m} )} 
\mathcal{F} ( T_m )
)_1
  \]
  as a consequence of $( \ast )$. Let $\sim$ be the congruence on
  $\mathcal{F}$($T_1 \boxtimes \cdots \boxtimes T_m$) generated by all
  non-contractible cubes in $T_1 \boxtimes \cdots \boxtimes T_m$. Then
  \[ ( C \circ \tmop{cosk}_{\Sigma} ) ( T_1 \boxtimes \cdots \boxtimes T_m )
     \cong ( C \circ \tmop{cosk} ) ( T_1 \boxtimes \cdots \boxtimes T_m )
     \cong \mathcal{F} ( T_1 \boxtimes \cdots \boxtimes T_m ) / \sim \]
  by $( \ast \ast )$ and corollary \ref{cor-rgrcubes}. But the
  non-contractible cubes in $T_1 \boxtimes \cdots \boxtimes T_m$ are the rigid
  ones, hence the map $\iota$ given on generators by
  \[ \begin{array}{llll}
       \iota : & ( ( \mathcal{C} \circ \tmop{cosk}_{\Sigma} )_1 &
\longrightarrow &
( 
\mathcal{F} ( T_1 ) 
\times_{\mathcal{F} (\Sigma_{1, 2} )}
  \mathcal{F} ( T_2 ) \times_{ \mathcal{F}
     ( \Sigma_{2, 3} )} \cdots 
\times_{\mathcal{F} ( \Sigma_{m - 1,m} )} 
\mathcal{F} ( T_m )
)_1\\
       &  ( e_1, \ldots, e_m ) & \longmapsto & ( ( e_1 ), \ldots, ( e_m ) )
     \end{array} \]
  is well defined. On the other hand, both categories have the same set of
  objects $\prod_{1 \leqslant i \leqslant m} ( T_i )_{_0}$ since the reflexive
  graphs $\Sigma_{i, i + 1}$  have one vertex for all $1 \leqslant i \leqslant
  m - 1$. Hence there is the functor
  \[ I : ( \mathcal{C} \circ \tmop{cosk}_{\Sigma} ) ( T_1 \boxtimes \cdots
     \boxtimes T_m ) \longrightarrow 
\mathcal{F} ( T_1 ) 
\times_{\mathcal{F} (\Sigma_{1, 2} )}
  \mathcal{F} ( T_2 ) \times_{ \mathcal{F}
     ( \Sigma_{2, 3} )} \cdots 
\times_{\mathcal{F} ( \Sigma_{m - 1,m} )} 
\mathcal{F} ( T_m )
\]
  with $I_0$ the identity map on $ \prod_{1 \leqslant i \leqslant m} ( T_i
  )_{_0}$  and $I_1 = \iota$. This functor has the obvious inverse.
\end{theproof}

Theorem \ref{theo-catsynchro} is not the most general statement of the kind,
the proof actually works for any wide pullback over reflexive graphs with
 one vertex. There could also be a ``labeled'' version, yet in our
applications we categorify precisely in order to get rid of the labels (it is
in the nature of a categorical semantics to be ``syntax free''). Philosophy
aside:

\begin{remark}
  \label{rem-cosk}Theorem \ref{theo-catsynchro} does not claim that
  \[ ( \mathcal{C} \circ \tmop{cosk}_{\Sigma} ) ( \Sigma_{t_i, t_{i + 1}} )
     \cong \mathcal{F} ( \Sigma_{t_i, t_{i + 1}} ) \]
  in which case $\mathcal{C} \circ \tmop{cosk}_{\Sigma}$ would preserve wide
  pullbacks of acubic reflexive graphs. Consider for instance the case $m =
  2$. What we do have is merely the diagram \\
  
  $
\xy
\xymatrix {(\mathcal{C} \:\circ\: \mathrm{cosk})(S \boxtimes T)
\ar[rrr] \ar[ddd] \ar[dr]^{\sim}_I
&&&
(\mathcal{C} \:\circ\: \mathrm{cosk})(T) 
\ar[ddd]_{!}
\ar@{=}[dl]_{\sim}
\\
&\mathcal{F}(S) \times_{\mathcal{F}(\Sigma_{s,t}} \mathcal{F}(S))
\ar[r] \ar[d]
&
\mathcal{F}(T) \ar[d]_{\mathcal{F}(j^T_{s,t}\:\circ\: t)}&
\\
& \mathcal{F}(S) \ar[r]_{\mathcal{F}(j^S_{s,t}\:\circ\: s)}
& \mathcal{F}(\Sigma_{s,t}) \ar[dr]^{!}&
\\
(\mathcal{C} \:\circ\: \mathrm{cosk}) (S) \ar@{=}[ur]^{\sim} 
\ar[rrr]_{!}
&&&
1
}
\POS(45,-20.5);\POS(45,-18.0) \connect@{-}
\POS(45,-20.5);\POS(42.5,-20.5) \connect@{-}
\endxy
$

 \mbox{}\\
by functoriality and remark \ref{rem-length}.
\end{remark}

\subsection{Change of Alphabet}

Suppose

\begin{center}
  $
\xymatrix {S \ar[r]^u \ar[d]_s & \hat{S} \ar[d]^{\hat{s}} \\
\Sigma \ar[r]_w & \hat{\Sigma}
}
$

\end{center}

{\noindent}and

\begin{center}
  $
\xymatrix {T \ar[r]^v \ar[d]_t & \hat{T} \ar[d]^{\hat{t}} \\
\Sigma \ar[r]_w & \hat{\Sigma}
}
$

\end{center}

{\noindent}commute. Let
\[ \begin{array}{llll}
     w_{\Upsilon} : & \Upsilon_{s, t} & \longrightarrow & \Upsilon_{\hat{s},
     \hat{t}}\\
     w_{s, \hat{s}} : & \Sigma_s & \longrightarrow & \hat{\Sigma}_{\hat{s}}\\
     w_{t, \hat{t}} : & \Sigma_t & \longrightarrow & \hat{\Sigma}_{\hat{t}}
   \end{array} \]
be the maps determined by $w$ and by universal property. ``Morally'', this
says that synchronization under this change of alphabet is still governed by
$\mathbf{s} (_-,_- )$. Let
\[ \begin{array}{llll}
     \pi_1 : & \Upsilon_{s, t} & \longrightarrow & \Sigma_s\\
     \pi_2 : & \Upsilon_{s, t} & \longrightarrow & \Sigma_t\\
     \hat{\pi}_1 : & \Upsilon_{\hat{s}, \hat{t}} & \longrightarrow &
     \hat{\Sigma}_{\hat{s}}\\
     \hat{\pi}_2 : & \Upsilon_{\hat{s}, \hat{t}} & \longrightarrow &
     \hat{\Sigma}_{\hat{t}}
   \end{array} \]
be the obvious projections. There is a commuting diagram

\begin{center}
  $
\xy
\xymatrix {& \Upsilon_{\hat{s},\hat{t}} \ar[rr]^{\hat{\pi}_2} 
\ar'[d][dd]^{\hat{\pi}_1} && 
{\hat{\Sigma}}_{\hat{t}} \ar[dd]
\\
\Upsilon_{s,t} \ar[rr]^>>>>>>{\pi_2} \ar[ru]^{w_\Upsilon} \ar[dd]_{\pi_1}
&& \Sigma_t \ar[ru]^{w_{t,\hat{t}}} \ar[dd] & 
\\
&{\hat{\Sigma}}_{\hat{s}}  \ar'[r][rr] && {\hat{\Sigma}}_{\hat{s},\hat{t}} 
\\
\Sigma_s \ar[rr] \ar[ru]^{w_{s,\hat{s}}} && \Sigma_{s,t} 
\ar@{.>}[ru]_{\bar{w}_{s,t}}
&
}
\POS(28,-37);\POS(28,-39.5) \connect@{-}
\POS(28,-37);\POS(30.5,-37) \connect@{-}
\POS(45,-23);\POS(45,-25.5) \connect@{-}
\POS(45,-23);\POS(47.5,-23) \connect@{-}

\endxy
$

\end{center}

{\noindent}with $\bar{w}$ given by universal property and thus

\begin{center}
  $
\xy
\xymatrix {& \hat{S} \boxtimes \hat{T} \ar[rr] \ar'[d][dd] && 
\hat{T} \ar[dd]
\\
S \boxtimes T \ar[rr] \ar@{.>}[ru]^{u \boxtimes v} \ar[dd]
&& T \ar[ru]^{v} \ar[dd] & 
\\
&\hat{S}  \ar'[r][rr] && {\hat{\Sigma}}_{\hat{s},\hat{t}} 
\\
S \ar[rr] \ar[ru]^{u} && \Sigma_{s,t} \ar[ru]_{\bar{w}_{s,t}}
&
}
\POS(6,-19);\POS(6,-16.5) \connect@{-}
\POS(6,-19);\POS(3.5,-19) \connect@{-}
\POS(26.5,-6);\POS(26.5,-3.5) \connect@{-}
\POS(26.5,-6);\POS(24.0,-6) \connect@{-}
\endxy
$

\end{center}

{\noindent}with $u \boxtimes v$ given by universal property. It is
straightforward that this fact generalizes to wide pullbacks, so theorem
\ref{theo-catsynchro} admits a ``relative'' version:

\begin{theorem}
  \label{theo-relative}Let  $w : \Sigma \rightarrow \hat{\Sigma}$ 
be a map and $(
  t_i : T_i \rightarrowlim \Sigma )_{1 \leqslant i \leqslant m}$ respectively
  $( \hat{t}_i : \hat{T}_i \rightarrowlim \hat{\Sigma} )_{1 \leqslant i
  \leqslant m}$ be morphisms of reflexive graphs with $T_i$ and 
  $\hat{T}_i$ acubic for all $1 \leqslant i \leqslant m$. Suppose there are
  morphisms $( u_i : T_i \rightarrowlim \hat{T}_i )_{1 \leqslant i \leqslant
  m}$ such that $\hat{t}_i \circ u_i = w \circ t_i$ for all $1 \leqslant i
  \leqslant m$. Let
  \[ \bar{w}_{i, i + 1} \equallim^{\tmop{def} .}  \bar{w}_{t_i, t_{i + 1}} :
     \Sigma_{t_i, t_{i + 1}} \longrightarrow 
\hat{\Sigma}_{\hat{t_i}, \hat{t}_{i + 1}} \]
  for all $1 \leqslant i \leqslant m - 1$. Then
  \[ ( \mathcal{C} \circ \tmop{cosk}_{\Sigma} ) ( u_1 \boxtimes \cdots
     \boxtimes u_m ) \cong \mathcal{F} ( u_1 ) \times_{\mathcal{F} (
     \bar{w}_{1, 2} ) \mathcal{}_{}}  \mathcal{F} ( u_2 ) \times_{\mathcal{}
     \mathcal{F} ( \bar{w}_{2, 3} )} \cdots \times_{\mathcal{} \mathcal{F} (
     \bar{w}_{m - 1, m} )} \mathcal{F} ( u_m ) \]
\end{theorem}

\begin{theproof}
  By chasing a series of diagrams as in remark \ref{rem-cosk}. The wide
  pullback
  \[ \mathcal{F} ( u_1 ) \times_{\mathcal{F} ( \bar{w}_{1, 2} )
     \mathcal{}_{}}  \mathcal{F} ( u_2 ) \times_{\mathcal{} \mathcal{F} (
     \bar{w}_{2, 3} )} \cdots \times_{\mathcal{} \mathcal{F} ( \bar{w}_{m - 1,
     m} )} \mathcal{F} ( u_m ) \]
  is in $\tmmathbf{\tmop{Cat}} ^{\rightarrow}$.
\end{theproof}

\begin{remark}
  \label{rem-ulf}The wide pullback
  \[ \mathcal{F} ( u_1 ) \times_{\mathcal{F} ( \bar{w}_{1, 2} )
     \mathcal{}_{}}  \mathcal{F} ( u_2 ) \times_{\mathcal{} \mathcal{F} (
     \bar{w}_{2, 3} )} \cdots \times_{\mathcal{} \mathcal{F} ( \bar{w}_{m - 1,
     m} )} \mathcal{F} ( u_m ) \]
  is actually in $\tmmathbf{\tmop{Ulf}}_{\mathcal{F}}^{\rightarrow}$ (c.f.
  section \ref{sec-2-thecorr}) since everything in sight is in the image of
  $\mathcal{F}$.
\end{remark}

\section{\label{sec-cip}The Programming Language {\tmname{CIP}}}

In this section we introduce the syntax and an operational semantics of the
programming language {\tmname{CIP}}. The acronym stands for ``communicating
imperative programs''. {\tmname{CIP}} is a variant of {\tmname{Concurrent
Pascal}} with {\tmname{CSP}}-style message-passing primitives. It is a nice
little language, as handy as an assembly language with a decent macro
expansion mechanism. Its syntax is presented by typing rules formulated with
respect to an underlying type theory $\mathcal{\mathcal{A}}$. We assume
$\mathcal{\mathcal{A}}$ is algebraic in order to fix the ideas and
keep the setup simple, yet more elaborated theories do also 
work \cite{bartbook}.
It's operational semantics is a set of rewrite rules over a data structure
consisting of instruction stacks, instruction registers and stores.

\subsection{{\tmname{CIP}}'s Syntax}

Let $\mathcal{V}$ be a countable set of {\tmem{variables}} and $\mathcal{P}$ a
countable set of {\tmem{port names}}. Let $\mathcal{A}$ be an algebraic theory
with $\mathcal{V}$ as set of variables.

\subsubsection{Grammar}

{\tmname{CIP}} consists of {\tmem{expressions}}, {\tmem{statements}} and
{\tmem{terms}}. {\tmname{CIP}}'s expressions are $\mathcal{A}$'s terms while
statements and terms are given by the productions
\begin{eqnarray*}
  \text{{\tmem{stat}}} & : \assign & \text{\texttt{nop}} \;|\; 
\text{\texttt{x:=}
  \texttt{}} e \;|\; \text{\texttt{p!}} \text{{\tmem{e}}} \;|\; 
\text{\texttt{p?x}}
  \;|\;\\
  &  & \text{{\tmem{stat}}}_1 ; \text{{\tmem{stat}}}_2 \;|\;\\
  &  & \mathtt{\tmop{if}} \; \text{{\tmem{e}}} \;
  \mathtt{\tmop{then}} \; \text{{\tmem{stat}}} \;
  \mathtt{\tmop{else}} \; \text{{\tmem{stat}}} \;
  \mathtt{\tmop{end}} \;|\;\\
  &  & \text{\texttt{while}}  \; \text{{\tmem{e}}} \;  \text{\texttt{do}} 
  \; \text{{\tmem{stat}}} \; \mathtt{\tmop{end}} \\
  \text{{\tmem{term}}} & : \assign & \text{{\tmem{stat}}} \;|\;
  \text{{\tmem{term}}} \ll \text{\texttt{p}}_1 \asymp \text{\texttt{q}}_1,
  \ldots, \text{\texttt{p}}_n \asymp \text{\texttt{q}}_n \gg
  \text{{\tmem{term}}}
\end{eqnarray*}
where {\tmem{e}} is an expression, $\text{\texttt{x}} \in \mathcal{V}$ , $n
\in \mathbbm{N}$ and $\text{\texttt{p}}_i, \text{\texttt{q}}_i \in
\mathcal{P}$ for all $0 \leqslant i \leqslant n$.

\subsubsection{\label{sec-3-typsequs}Typing Sequents}

Let $\mathbbm{T}_{\mathcal{A}}$ be $\mathcal{A}$'s set of types. The general
format of a typing sequent for {\tmname{CIP}} is
\[ \text{\texttt{x}}_1 : \tau_1, \ldots, \text{\texttt{x}}_n : \tau_n
   \hspace{0.75em} \vdash \hspace{0.75em} \text{{\tmem{term}}} :
   \hspace{0.5em} \left\langle \text{\texttt{p}}_1 : \theta_1^{s_1}, \ldots,
   \text{\texttt{p}}_m : \theta_m^{s_m} \right\rangle \]
where
\begin{itemizeminus}
  \item $( \text{\texttt{x}} \text{\texttt{}}_i, \tau_i ) \in \mathcal{V}
  \times \mathbbm{T}_{\mathcal{A}}$ for all $1 \leq i \leq n$
  
  \item $( \text{\texttt{p}}_j, \theta_j ) \in \mathcal{P} \times
  \mathcal{\mathbbm{T}_{\mathcal{A}}}$ for all $1 \leq j \leq m$
  
  \item $s_j \in \{ -, + \}$ for all $1 \leq j \leq m$
\end{itemizeminus}
The $x_i$'s are names of variables while the entries $p_k : \theta_k^{s_k}$
are pairs consisting of a port name $p_k$ designating an interface port of
type $\theta_k$. The latter is decorated with a {\tmem{polarity}}
$s_k$ which discriminates if $p_k$ is an input or an output port. Their list
represents the \textit{signature} of the program as seen by its
\textit{environment}.

A {\tmem{well-formed context}} is a context without a repetition of a
variable name. A {\tmem{well-formed signature}} is a signature without a
repetition of a port name. In other words, well-formed contexts and signatures
encode partial functions of types $\mathcal{V} \rightharpoonup
\mathbbm{T}_{\mathcal{A}}$ respectively $\mathcal{P} \rightharpoonup
\mathbbm{T}_{\mathcal{A}} \times \text{\{ -,+ \}}$. Let
$\tmmathbf{\mathcal{K}}$ be the set of all well-formed contexts and
$\tmmathbf{\mathcal{S}}$ be the set of all well-formed signatures.

\subsubsection{Typing Rules}

We assume well-formed all the contexts and signatures occuring in the premises
of the following typing rules.

\paragraf{Structural Rules} Let $S_n$ be the $n$th symmetric group.
{\tmname{Cip}}'s structural rules are
\[ \text{\textbf{{\tmname{Weak}}} $\frac{\Gamma \hspace{0.25em} \vdash
   \hspace{0.25em} \text{{\tmem{term}}} \hspace{0.25em} : \hspace{0.25em}
   \Delta}{\Gamma, \hspace{0.25em} x : \tau \hspace{0.25em} \vdash
   \hspace{0.25em} \text{{\tmem{term}}} \hspace{0.25em} : \hspace{0.25em}
   \Delta}$}  \;\;\; 
\text{\textbf{{\tmname{Perm}}} $\frac{\Gamma \hspace{0.25em}
   \vdash \hspace{0.25em} \text{{\tmem{term}}} \hspace{0.25em} :
   \hspace{0.25em} \Delta \pi_v \in S_{\text{} | \Gamma |} \pi_s \in S_{|
   \Delta |}}{\pi_v \Gamma \hspace{0.25em} \vdash \hspace{0.25em}
   \text{{\tmem{term}}} \hspace{0.25em} : \hspace{0.25em} \pi_s \Delta}$} \]
\paragraf{Sequential Fragment} {\tmname{CIP}}'s sequential fragment is given
by the rules
\[ \text{\textbf{{\tmname{Asg}}} $\frac{\Gamma, \hspace{0.25em} \mathtt{x} :
   \tau \hspace{0.25em} \vdash \hspace{0.25em} e : \hspace{0.25em}
   \tau}{\Gamma, \mathtt{x} : \tau \hspace{0.25em} \vdash \hspace{0.25em}
   \mathtt{x} : = e : \hspace{0.25em} \left\langle \hspace{0.25em}
   \right\rangle}$}  \;\;\; \text{ \textbf{{\tmname{Nop}}}
   $\frac{}{\mathtt{\tmop{nop}} : \hspace{0.25em} \left\langle \hspace{0.25em}
   \right\rangle}$} \]
\[ \text{{\tmname{\textbf{If}}} $\frac{\Gamma \hspace{0.25em} \vdash
   \hspace{0.25em} e : \hspace{0.25em} \mathtt{\tmop{bool}} \quad \Gamma
   \hspace{0.25em} \vdash \hspace{0.25em} \text{{\tmem{stat}}}_1 :
   \hspace{0.25em} \Delta \quad \Gamma \hspace{0.25em} \vdash \hspace{0.25em}
   \text{{\tmem{stat}}}_2 : \hspace{0.25em} \Theta}{\Gamma \hspace{0.25em}
   \vdash \hspace{0.25em} \mathtt{\tmop{if}} \hspace{0.5em} e \hspace{0.5em}
   \mathtt{\tmop{then}} \hspace{0.5em} \text{{\tmem{stat}}}_1 \hspace{0.5em}
   \mathtt{\tmop{else}} \hspace{0.5em} \text{{\tmem{stat}}}_2 \hspace{0.5em}
   \mathtt{\tmop{end}} : \hspace{0.25em} \Delta \otimes \Theta}$} \]

\[ \text{{\tmname{\textbf{While}}}$\frac{\Gamma \hspace{0.25em} \vdash
    \hspace{0.25em} e : \hspace{0.25em} \mathtt{\tmop{bool}} \quad \Gamma
    \hspace{0.25em} \vdash \hspace{0.25em} \text{{\tmem{stat}}} :
    \hspace{0.25em} \Delta}{\Gamma \hspace{0.25em} \vdash \hspace{0.25em}
    \mathtt{\tmop{while}}  \hspace{0.5em} e \hspace{0.5em} \mathtt{\tmop{do}}
    \hspace{0.5em} \text{{\tmem{stat}}} \hspace{0.5em} \mathtt{\tmop{end}} :
    \hspace{0.25em} \Delta}$}\]  
\[ \text{\textbf{{\tmname{Seq}}} $\frac{\Gamma \hspace{0.25em} \vdash
   \hspace{0.25em} \text{{\tmem{stat}}}_1 : \hspace{0.25em} \Delta \quad
   \Gamma \hspace{0.25em} \vdash \hspace{0.25em} \text{{\tmem{stat}}}_2 :
   \hspace{0.25em} \Theta}{\Gamma \hspace{0.25em} \vdash \hspace{0.25em}
   \text{{\tmem{stat}}}_1 ; \text{{\tmem{stat}}}_2 : \hspace{0.25em} \Delta
   \otimes \Theta}$} \]
where $\tau_i \in \mathcal{\mathbbm{T}_{\mathcal{A}}}$ for all $i \in
\mathbbm{N}$ while $\Delta \otimes \Theta \in \tmmathbf{\mathcal{S}}$ is the
concatenation of $\Delta$ and $\Theta$, possibly after  $\alpha$-conversion.

\paragraf{Concurrent Fragment} {\tmname{CIP}}'s concurrent fragment is given
by the input/output rules
\[ \text{\textbf{{\tmname{Out}}} $\frac{\Gamma \hspace{0.25em} \vdash
   \hspace{0.25em} e : \tau}{\Gamma \hspace{0.25em} \vdash \hspace{0.25em}
   \mathtt{p} !e : \hspace{0.25em} \left\langle \mathtt{p} : \hspace{0.25em}
   \tau^{\hspace{0.25em} +} \right\rangle}$}  \;\;\;\; \text{{\tmname{\textbf{In}}}
   $\frac{}{\Gamma, \mathtt{x} : \tau \hspace{0.25em} \vdash \hspace{0.25em}
   \mathtt{p} ? \mathtt{x} : \hspace{0.25em} \left\langle \mathtt{p} :
   \hspace{0.25em} \tau^{\hspace{0.25em} -} \right\rangle}$} \]
\paragraf{Composition rule} {\tmname{CIP}}'s composition is given by the
rule
\[ 
\frac{
  \begin{array}{c}
     \Gamma \; \vdash \; 
     \text{\tmem{term}}_1 \; : \; \Delta, \;
     \left\langle \mathtt{p}_1 : \; \tau_1^{\; +},
     \ldots, \mathtt{p}_n : \; \tau_n^{\; +}
     \right\rangle 
     \\
     \Xi \; \vdash \; 
     \text{\tmem{term}}_2 \; : \; \Theta, \;
     \left\langle \mathtt{q}_1 : \; \tau_1^{\; -},
     \ldots, \mathtt{q}_n : \; \tau_n^{\; -}
     \right\rangle 
   \end{array}}
   {\Gamma \oplus \Xi \; \vdash \;
   \text{\tmem{term}}_1 \; \ll \mathtt{p}_1 \asymp
   \mathtt{q}_1, \ldots, \mathtt{p}_n \asymp \mathtt{q}_n
   \gg \; \text{\tmem{term}}_2 \; :
   \; \Delta \otimes \Theta}
\]
where $\Gamma \oplus \Xi \in \tmmathbf{\mathcal{K}}$ is the concatenation (of
both components) of $\Gamma$ and $\Xi$, possibly after  $\alpha$-conversion.

A pair $\text{\texttt{p}}_i \asymp \text{\texttt{}} \text{\texttt{q}}_i$ is a
{\tmem{restricted channel}} of type $\tau_i$, connecting ports
$\text{\texttt{p}}_i$ and $\text{\texttt{q}}_i$. The notation resembles the
one used in linear logic since it is where the inspiration comes from, just
think of the interface ports as resources. In more practical terms,
{\tmstrong{{\tmname{MCut}}}} allows to define interactions of processes
connected by typed channels.

\subsubsection{\label{sec-3-equterm}Equality on Terms} 

We are in fact more interested in the ``end product'' of a series of
applications of the typing rules than in order the latter were performed.
Order-sensitive rules are {\tmname{{\tmstrong{Seq}}}} and
{\tmname{{\tmstrong{MCut}}}}, applications of the latter possibly requiring
$\alpha$-conversion. Given $\Gamma \in \tmmathbf{\mathcal{K}}$ and $\Delta \in
\tmmathbf{\mathcal{S}}$, let $\mathcal{T}_{_{\Gamma, \Delta}}$ be the set of
well-formed terms $\Gamma \vdash t : \Delta$. The equivalence relation
$\approx \in^{} \mathcal{T}_{_{\Gamma, \Delta}} \times \mathcal{T}_{_{\Gamma,
\Delta}}$ is generated by
\begin{enumerateroman}
  \item $( t ; t' ) ; t'' \approx t ; ( t' ; t'' )$;
  
  \item $\left( t \ll L \gg t' \right) \ll K \gg t'' \approx t \ll L \gg_{} 
  \left( t' \ll K \gg t''  \right) $
\end{enumerateroman}
In particular, we can write $t ; t', t''$ and  $t \ll L \gg t' \ll K \gg t''$
without ambiguity. Let $\mathcal{L}_{_{\Gamma, \Delta}} \subseteq
\mathcal{T}_{_{\Gamma, \Delta}}$ be $\mathcal{T}_{_{\Gamma, \Delta}}$'s subset
of  well-formed {\tmem{statements}}. The equivalence class $[ t ]$ of $t \in
\mathcal{T}_{_{\Gamma, \Delta}}$ can be characterized by a pair
\[ \left( \left[ \text{{\tmem{s}}}_1, \ldots, \text{{\tmem{s}}}_n ] \right., [
   \text{\texttt{p}}_1 \asymp \text{\texttt{q}}_1, \ldots, \text{\texttt{}}
   \text{\texttt{}} \text{\texttt{q}}_m \asymp \text{\texttt{q}}_m ] ) \right.
\]
for some $n, m \in \mathbbm{N}$, such that $\text{{\tmem{s}}}_i \in
\mathcal{L}_{_{\Gamma, \Delta}}$ for all $i \in \mathbbm{N}$. All the
(sub)statements are flattened with respect to the sequencing operator ``;''.
Such a pair, called $t$'s {\tmem{normal form}}, is unique up to
$\alpha$-conversion.

\begin{proposition}
  \label{prop-nform} Let $\Gamma \hspace{0.25em} \vdash \hspace{0.25em}
  \text{{\tmem{t}}} \hspace{0.25em} : \hspace{0.25em} \Delta$ be a
term. The following are equivalent
  \begin{enumerateroman}
    \item $t$'s normal form is $\left( \left[ \text{{\tmem{s}}}_1, \ldots,
    \text{{\tmem{s}}}_n ] \right., [ \text{\texttt{p}}_1 \asymp
    \text{\texttt{q}}_1, \ldots, \text{\texttt{}} \text{\texttt{}}
    \text{\texttt{q}}_m \asymp \text{\texttt{q}}_m ] ) \right.$;
    
    \item there is a family of statements $( \Gamma \hspace{0.25em}_i \vdash
    \hspace{0.25em} \text{{\tmem{s}}}_i \hspace{0.25em} : \hspace{0.25em}
    \Delta_i )_{1 \leqslant i \leqslant n}$ such that
    \[ \left( \Gamma \hspace{0.25em} \vdash \hspace{0.25em} \text{{\tmem{t}}}
       \hspace{0.25em} : \hspace{0.25em} \Delta \right) = \left( \Gamma_1
       \oplus \cdots \oplus \Gamma_n \vdash s_1 \ll \Phi_{1, 2} \gg s_2 \cdots
       s_{n - 1} \ll \Phi_{n - 1, n} \gg s_n : \Delta'_1 \otimes \cdots
       \otimes \Delta_n' \right) \]
    with $\Delta'_1, \cdots, \Delta_n' = ( \Delta_1, \cdots \Delta_n )
    \backslash [ \mathtt{p}_1 : \hspace{0.25em} \tau_1^{\hspace{0.25em} +},
    \ldots, \mathtt{p}_n : \hspace{0.25em} \tau_n^{\hspace{0.25em} +},
    \mathtt{q}_1 : \hspace{0.25em} \tau_1^{\hspace{0.25em} -}, \ldots,
    \mathtt{q}_n : \hspace{0.25em} \tau_n^{\hspace{0.25em} -} ]$ and $[
    \Phi_{1, 2}, \ldots, \Phi_{n - 1, n} ] = [ \text{\texttt{p}}_1 \asymp
    \text{\texttt{q}}_1, \ldots, \text{\texttt{p}}_n \asymp
    \text{\texttt{q}}_n ]$ up to $\alpha$-conversion.
  \end{enumerateroman}
\end{proposition}

\begin{theproof}
  Obvious.
\end{theproof}

{\noindent}Let
\begin{eqnarray*}
  \mathcal{T}_{_{\Gamma}} & \equallim^{\tmop{def} .} & \bigcup_{\Delta \in
  \tmmathbf{\mathcal{S}}} \mathcal{T}_{_{\Gamma, \Delta}}\\
  \mathcal{T} & \equallim^{\tmop{def}} & \bigcup_{\Gamma \in
  \tmmathbf{\mathcal{K}}} \mathcal{T}_{_{\Gamma}}
\end{eqnarray*}
and $\mathcal{L}_{_{\Gamma}} \subseteq \mathcal{T}_{_{\Gamma}}$ and
$\mathcal{L} \subseteq \mathcal{T}$ be the corresponding subsets of
well-formed statements.

\subsection{\label{sec-2-opsem} {\tmname{CIP}}'s Operational Semantics}

{\tmname{CIP}}'s operational semantics is given in continuation style, i.e. by
an abstract machine.

\subsubsection{\label{sec-3-design}The Abstract Machine}

\paragraf{Stores} Let $\left\lceil \_ \right\rceil$ be an interpretation of
$\mathcal{A}$ in $\mathbf{\tmop{Set}}$ and
\[ \Theta_{\mathcal{A}} \equallim^{\tmop{def} .} \bigcup_{\tau \in
   \mathcal{\mathbbm{T}}_{\mathcal{A}}} \left\lceil \tau \right\rceil \]
A \textit{store} $\sigma$ is a partial function $\sigma \in \mathcal{V}
\rightharpoonup \Theta_{\mathcal{A}}$ with finite domain of definition. A
$\Gamma$-store is a store defined on $\text{Var} \left( \Gamma \right)$ such
that
\[ \forall \left( x : \tau \right) \in \Gamma . \hspace{0.25em} \sigma \left(
   x \right) \in \left\lceil \tau \right\rceil \]
We write $\mathcal{S_{\Gamma}}$ for their set.

\paragraf{Locations} A $\Gamma$-{\tmem{location}} is the address of a node in
the syntax tree of some $\text{{\tmem{stat}}} \in \mathcal{L}_{_{\Gamma}}$.
Let $L_{_{\Gamma}}$ be the set of the $\Gamma$-locations under some encoding,
e.g. a path in the tree. Let $\rho ( \text{{\tmem{stat}}} )$ be the root of
the corresponding tree. Let further
\[ \eta, \xi_1, \xi_2, \phi :\mathcal{L}_{_{\Gamma}} \times L_{_{\Gamma}}
   \rightharpoonup L_{_{\Gamma}} \]
be the partial operations such that
\begin{itemizeminus}
  \item $\xi_1 ( \mathtt{\tmop{if}} \hspace{0.5em} e \hspace{0.5em}
  \mathtt{\tmop{then}} \hspace{0.5em} \text{{\tmem{stat}}}_1 \hspace{0.5em}
  \mathtt{\tmop{else}} \hspace{0.5em} \text{{\tmem{stat}}}_2 \hspace{0.5em}
  \mathtt{\tmop{end}}, l )$
returns the location of $\text{{\tmem{stat}}}_1$
  given the location $l$ of the if-statement
  
  \item $\xi_1 ( \mathtt{\tmop{if}} \hspace{0.5em} e \hspace{0.5em}
  \mathtt{\tmop{then}} \hspace{0.5em} \text{{\tmem{stat}}}_1 \hspace{0.5em}
  \mathtt{\tmop{else}} \hspace{0.5em} \text{{\tmem{stat}}}_2 \hspace{0.5em}
  \mathtt{\tmop{end}}, l )$returns the location of $\text{{\tmem{stat}}}_2$
  given the location $l$ of the if-statement
  
  \item $\phi ( \mathtt{\tmop{while}} e \hspace{0.5em} \mathtt{\tmop{do}}
  \hspace{0.5em} \text{{\tmem{stat}}} \hspace{0.5em} \mathtt{\tmop{end}}, l )$
  returns the location of $\text{{\tmem{stat}}}$ given the location $l$ of the
  while-statement
\end{itemizeminus}
We write $\xi_1 ( l )$, $\xi_2 ( l )$ respectively $\phi ( l )$ when it is
clear from the context what the statements are.

\paragraf{Processes} Given $\Gamma \in \tmmathbf{\mathcal{K}}$ let
$\mathcal{M}_{_{\Gamma}}$ be the set of $\Gamma$-stores. A
$\Gamma$-{\tmem{process}} is a triple
\[ ( S, \langle \text{{\tmem{stat}}}, l \left\rangle, \sigma \right) \in
   \text{List} (\mathcal{L}_{{\Gamma}}
\times L_{_{\Gamma}} )
   \times (\mathcal{L}_{_{\Gamma}} \times L_{_{\Gamma}} )_{\text{{\small
   {\footnotesize }}}} \times \mathcal{M}_{_{\Gamma}} \]
$S$ is the instruction stack while {\tmem{stat}} is the instruction register.
The reason to tie the instructions to their locations will become apparent in
section \ref{sec-evo}.

\paragraf{Configurations and Rewrites} A $\Gamma$-{\tmem{configuration}}
{\tmem{}} is a vector of $\Gamma$-processes. {\tmname{CIP}}'s operational
semantics consists of a set of conditional rewrite rules
\[ \frac{\varpi}{[ P_1, \ldots, P_n ] \longrightarrow [ P_1', \ldots, P_n' ]}
\]
on $\Gamma$-configurations. We consider fixed {\tmem{communication
topologies}}, i.e. $n$ remains constant under the rewrites. The vector of
channels, $\Gamma$ itself as well as $\left\lceil \_ \right\rceil$ will be
needed to evaluate the rewrite conditions $\varpi$.

Observe that, although we need $\Gamma$ for rewrite conditions, we can keep
it as a constant datum much like the list of channels. It is a design choice
since we do not introduce nested scopes or (remote) procedure calls, a 
side-issue here. In presence such features though, rewrites may
not leave $\Gamma$ constant.

\subsubsection{\label{sec-3-seqfrag}Rewrite Rules for {\tmname{cip}}'s
{\tmname{}}Sequential Fragment}

Let $f$ and $f'$ be partial functions and
\[ \left( f \uplus f' \right) \left( x \right) \equallim^{\tmop{def}.}
   \left\{ \begin{array}{cc}
     f \left( x \right) & x \in \text{dom} \left( f \right) \setminus
     \text{dom} \left( f' \right)\\
     f' \left( x \right) & x \in \text{dom} \left( f' \right)\\
     \uparrow & 
   \end{array} \right. \]

Given a well-formed expression $\Gamma \hspace{0.25em} \vdash \hspace{0.25em}
e : \tau$ such that $[ x_1 : \tau_1, \ldots, x_n : \tau_n ] \subseteq
\Gamma_0$ and $\sigma$ a $\Gamma$-store, let further
\[ \sigma \left( e \right) \equallim^{\tmop{def} .} \left\lceil \Gamma
   \hspace{0.25em} \vdash \hspace{0.25em} e \right\rceil \left( \sigma \left(
   x_1 \right) \ldots \sigma \left( x_n \right) \right) \in \left\lceil \tau
   \right\rceil \]
\begin{flushleft}
  {\tmname{CIP}}'s sequential fragment is given by the rather self-explaining
  rules
\end{flushleft}
\[ \text{{\tmstrong{{\tmname{Nop}}}}}  \frac{}{\begin{array}{l}
     \left[ P_1, \ldots, P_{i - 1}, \left( S \ast \left[ \langle
     \text{{\tmem{stat}}}, l \rangle \right], \langle \mathtt{\tmop{nop}}, k
     \rangle, \sigma \right), \ldots, P_n \right] \rightarrow\\
     \left[ P_1, \ldots, P_{i - 1}, \left( S, \langle \text{{\tmem{stat}}}, l
     \rangle, \sigma \right), \ldots, P_n \right]
   \end{array}} \]
\[ \text{{\tmstrong{{\tmname{Asg}}}}}  \frac{}{\begin{array}{l}
     \left[ P_1, \ldots, P_{i - 1}, \left( S \ast \left[ \langle
     \text{{\tmem{stat}}}, l \rangle \right], \langle \mathtt{x} : = e, k
     \rangle, \sigma \right), \ldots, P_n \right] \rightarrow\\
     \left[ P_1, \ldots, P_{i - 1}, \left( S, \langle \text{{\tmem{stat}}}, l
     \rangle, \sigma \uplus \left[ \mathtt{x} \mapsto \sigma \left( e \right)
     \right] \right), \ldots, P_n \right]
   \end{array}} \]
\[ \text{{\tmstrong{{\tmname{If1}}}}}  \frac{\sigma \left( e \right)
   \;=\; t  \! t}{\begin{array}{l}
     \left[ P_1, \ldots, P_{i - 1}, \left( S, \langle \mathtt{\tmop{if}} 
     \hspace{0.5em} e \hspace{0.5em} \mathtt{\tmop{then}} \hspace{0.5em}
     \text{{\tmem{stat}}}_1 \hspace{0.5em} \mathtt{\tmop{else}} \hspace{0.5em}
     \text{{\tmem{stat}}}_2 \hspace{0.5em} \mathtt{\tmop{end}}, l \rangle,
     \sigma \right), \ldots, P_n \right] \rightarrow\\
     \left[ P_1, \ldots, P_{i - 1}, \left( S, \langle \text{{\tmem{stat}}}_1,
     \xi_1 ( l ) \rangle, \sigma \right), \ldots, P_n \right]
   \end{array}} \]
\[ \text{{\tmstrong{{\tmname{If2}}}}}  \frac{\sigma \left( e \right)
   \;=\; f \!\! f}{\begin{array}{l}
     \left[ P_1, \ldots, P_{i - 1}, \left( S, \langle \mathtt{\tmop{if}} e
     \hspace{0.5em} \mathtt{\tmop{then}} \hspace{0.5em} \text{{\tmem{stat}}}_1
     \hspace{0.5em} \mathtt{\tmop{else}} \hspace{0.5em} \text{{\tmem{stat}}}_2
     \hspace{0.5em} \mathtt{\tmop{end}}, l \rangle, \sigma \right), \ldots,
     P_n \right] \rightarrow\\
     \left[ P_1, \ldots, P_i, \left( S, \langle \text{{\tmem{stat}}}_2, \xi_2
     ( l ) \rangle, \sigma \right), \ldots, P_n \right]
   \end{array}} \]
\[ \text{{\tmstrong{{\tmname{While1}}}}}  \hspace{0.25em} \frac{\sigma \left(
   e \right) \;=\; t \! t}{\begin{array}{l}
     \left[ P_1, \ldots, P_{i - 1}, \left( S, \langle \mathtt{\tmop{while}}
     \hspace{0.5em} e \hspace{0.5em} \mathtt{\tmop{do}} \hspace{0.5em}
     \text{{\tmem{stat}}} \hspace{0.5em} \mathtt{\tmop{end}}, l \rangle,
     \sigma \right), \ldots, P_n \right] \rightarrow\\
     \left[ P_1, \ldots, P_{i - 1}, \left( S \ast \left[ \langle
     \mathtt{\tmop{while}} e \hspace{0.5em} \mathtt{\tmop{do}} \hspace{0.5em}
     \text{{\tmem{stat}}} \hspace{0.5em} \mathtt{\tmop{end}}, l \rangle
     \right], \langle \text{{\tmem{stat}}}, \phi ( l ) \rangle, \sigma
     \right), \ldots, P_n \right]
   \end{array}} \]
\[ \text{{\tmstrong{{\tmname{While2}}}}}  \frac{\sigma \left( e \right)
   \;=\; f \! \! f}{\begin{array}{l}
     \left[ P_1, \ldots, P_{i - 1}, \left( S \ast \left[ \langle
     \text{{\tmem{stat}}}_1, l_1 \rangle \right], \langle
     \mathtt{\tmop{while}} \hspace{0.5em} e \hspace{0.5em} \mathtt{\tmop{do}}
     \hspace{0.5em} \text{{\tmem{stat}}} \hspace{0.5em} \mathtt{\tmop{end}}, l
     \rangle, \sigma \right), \ldots, P_n \right] \rightarrow\\
     \left[ P_1, \ldots, P_{i - 1}, \left( S, \langle \text{{\tmem{stat}}}_1,
     l_1 \rangle, \sigma \right), \ldots, P_n \right]
   \end{array}} \]
There is no explicit rewriting rule corresponding to the sequence operator
``;''. Such a rule, e.g
\[ \frac{}{\begin{array}{l}
     \left[ P_1, \ldots, P_{i - 1}, ( S, \langle \text{{\tmem{stat}}}
     \text{}_1 ; \text{{\tmem{stat}}}_2, l \rangle, \sigma ), \ldots, P_n
     \right] \rightarrow\\
     \left[ P_1, \ldots, P_{i - 1}, ( S \ast \left[ \langle
     \text{{\tmem{stat}}}_2, \eta ( l ) \rangle \right], \langle
     \text{{\tmem{stat}}}_1, l \rangle, \sigma ), \ldots, P_n \right]
   \end{array}} \]
(where $\text{} \eta ( \text{{\tmem{stat}}}_1 ; \text{{\tmem{stat}}}_2, l )$
returns the location of $\text{{\tmem{stat}}}_2$ given the location $l$ of
$\text{\tmem{stat}}_1$ involves only the management of the instruction stack.
We therefore assume the rule implicit to the abstract machine under
consideration (``hardwired'' to use a real-worldish jargon).

It is intuitively clear that transitions obtained by the sequential rules
shall not be observable. 

\subsubsection{\label{sec-3-concfrag}Rewrite Rules for {\tmname{cip}}'s
{\tmname{}}Concurrent fragment}

CIP's message-passing mechanism is given by the rules detailed as follows.

\paragraf{Rendez-vous} The ``rendez-vous'' rule
\[ \text{{\tmstrong{RV}}} \hspace{0.25em}  \frac{\mathtt{p} \asymp
   \mathtt{q}}{\left[ \begin{array}{l}
     P_1\\
     \vdots\\
     P_i\\
     \left( S_1 \ast \left[ \langle \text{{\tmem{stat}}}_1, l_1 \rangle
     \right], \langle \mathtt{p!} e, l \rangle, \sigma_1 \right)\\
     \vdots\\
     P_{i + m}\\
     \left( S_2 \ast \left[ \langle \text{{\tmem{stat}}}_2, k_2 \rangle
     \right], \langle \mathtt{q} ? \mathtt{x}, k \rangle, \sigma_2 \right)\\
     \vdots\\
     P_n
   \end{array} \right] \rightarrow \left[ \begin{array}{l}
     P_1\\
     \vdots\\
     P_i\\
     ( S_1, \langle \text{{\tmem{stat}}}_1, l_1 \rangle, \sigma_1 )\\
     \vdots\\
     P_{i + m}\\
     \left( S_2, \langle \text{{\tmem{stat}}}_2, k_2 \rangle, \sigma_{i + m}
     \uplus \left[ \mathtt{x} \mapsto \sigma_i \left( e \right) \right]
     \right)\\
     \vdots\\
     P_n
   \end{array} \right]} \]
specifies the transmission of a datum. Process $P_{i + 1}$ sends the
expression $e$ through the port \texttt{p} and proceeds by popping the next
statement from the stack. On the other side, process $P_{i + m + 1}$ is ready
to receive a datum from the port \texttt{q}. Both ports are connected, i.e.
they form a restricted channel $\text{\texttt{p}} \asymp \text{\texttt{q}}$,
which allows $P_{i + m + 1}$ to receive the expression $e$ evaluated in its
original context. $P_{i + m + 1}$ stores it in the variable \texttt{x} and
proceeds by popping the next statement from the stack. Transitions obtained by
{\tmname{{\tmstrong{RV}}}} shall not be observable since the communication
channel is restricted.

\paragraf{Send} The ``send'' rule

\[ \text{{\tmstrong{S}}} \hspace{0.25em}  \frac{\nexists \hspace{0.25em}
   \mathtt{q} . \hspace{0.25em} \hspace{0.25em} \mathtt{p} \asymp
   \mathtt{q}}{\begin{array}{l}
     \begin{array}{l}
       \left[ P_1, \ldots, P_{i - 1}, \left( S \ast \left[ \langle
       \text{{\tmem{stat ,l}}} \rangle \text{{\tmem{}}} \right], \langle
       \mathtt{p!} e, k \rangle, \sigma \right), \ldots, P_n \right]
     \end{array} \rightarrow\\
     \begin{array}{l}
       \left[ P_1, \ldots, P_{i - 1}, \left( S, \langle \text{{\tmem{stat
       ,l}}} \rangle, \sigma \right), \ldots, P_n \right]
     \end{array}
   \end{array}} \]
specifies the behavior of a program where process $P_i$ performs a ``send''
without a rendez-vous partner.  Basically nothing happens, yet a transition
obtained by an application of {\tmname{{\tmstrong{S}}}} shall be observable.

\paragraf{Receive} The ``receive'' rules
\[ \text{{\tmstrong{R}}} \hspace{0.25em}_w  
\frac{\nexists \hspace{0.25em}
   \mathtt{p} . \hspace{0.25em}  \hspace{0.25em} \mathtt{p} \asymp \mathtt{q}
    \hspace{1.5em} w \in \left\lceil \Gamma ( x ) \right\rceil}
{\begin{array}{l}
     \begin{array}{l}
       \left[ P_1, \ldots, P_{i - 1}, \left( S \ast \left[ \langle
       \text{{\tmem{stat ,l}}} \rangle \right], \langle \mathtt{q} ?
       \mathtt{x}, k \rangle, \sigma \right), \ldots, P_n \right]
     \end{array} \rightarrow\\
     \begin{array}{l}
       \left[ P_1, \ldots, P_{i - 1}, \left( S, \langle \text{{\tmem{stat
       ,l}}} \rangle \text{{\tmem{}}}, \sigma \uplus [ \text{\texttt{x}}
       \longmapsto w ] ), \ldots, P_n \right] \right.
     \end{array}
   \end{array}} \]
specify the behavior of a program where process $P_i$ performs a ``receive''
without a rendez-vous partner. They collectively say that the variable
\texttt{x} can then be assigned to any value $w$ of the right type.
Transitions obtained applications of the {\tmname{{\tmstrong{R}}}} $_w$'s
shall be observable.

\subsection{State Space}

\begin{definition}
  \label{def-confs}Let $t \in \mathcal{T}_{_{\Gamma}}$  with normal form
\[
\left(  
[s_1,\ldots,s_n],[\mathtt{p}_1 \asymp \mathtt{q}_1,\ldots
\mathtt{p}_m \asymp \mathtt{q}_m]
\right)
\]
  (c.f. \ref{sec-3-equterm}) and $\mu \equallim^{\tmop{def} .} [ \mu_1,
  \ldots, \mu_n ]$ be a vector of $\Gamma$-stores (c.f. \ref{sec-3-design}).
  Let
  \[  t_{\mu} \;\equallim^{\tmop{def} .}\; \left[ ( [\: ], \langle
     \text{{\tmem{s}}}_1, \text{} \rho ( \text{{\tmem{s}}}_1 ) \rangle, \mu_1
     ), \ldots, ( [\: ], \langle \text{{\tmem{s}}}_n, 
\rho ( \text{{\tmem{s}}}_n
     ) \rangle, \mu_n ) \right] \]
  be a $\Gamma$configuration. A $( \Gamma, t, \mu )$-configuration {\tmem{}}is
  $t_{\iota}$ or is obtained from $t_{\iota}$ by applications of the rewrite
  rules of section \ref{sec-3-seqfrag} and section \ref{sec-3-concfrag}. Let
  $\mathcal{R}_{t, \mu}$ be the set of all  $( \Gamma, t, \mu
  )$-configurations. The state space of $t$ is defined as
  \[ \Psi_t \equallim^{\tmop{def} .}  \bigcup_{\mu \in \mathcal{S_{\Gamma}}}
     \mathcal{R}_{t, \mu} \]
\end{definition}

\begin{proposition}
  \label{prop-storemap}Under the notation of definition \ref{def-confs}, there
  is a unique
  \[ \begin{array}{llll}
       |_- | : & \Psi_t & \longrightarrow & \prod_{1
       \leqslant i \leqslant n} \left\lceil \tau_i \right\rceil\\
       &  &  & \\
       & \left[ \begin{array}{l}
         P_1\\
         \vdots\\
         P_n
       \end{array} \right] & \longmapsto & \left[ \sigma_1 ( x_1^{{\small ( 1
       )}} ), \ldots, \sigma_1^{{\small }} ( x_1^{{\small ( m_1 )}} ), \ldots,
       \sigma_n ( x_n^{{\small ( 1 )}} ), \ldots, \sigma_n ( x_n^{( m_n )} )
       \right]
     \end{array} \]
  such that $\sigma_j \in \mathcal{S}_{\Gamma}$ is minimal for $1 \leqslant j
  \leqslant n$.
\end{proposition}

\begin{theproof}
  Let $\Gamma = x_1 : \tau_1, \ldots,_{} x_k : \tau_k$. By proposition
  \ref{prop-nform} we can assume that there is a unique decomposition
  \[ x_1 : \tau_1, \ldots,_{} x_k : \tau_k = \left( x_i^{{\small ( 1 )}} :
     \tau_i^{( 1 )}, \ldots, x_i^{{\small ( m_i )}} : \tau_i^{( m_i )}
     \right)_{1 \leqslant i \leqslant k} \]
  such that each $\sigma_i$ is a $( x_i^{{\small ( 1 )}} : \tau_i^{( 1 )},
  \ldots, x_i^{{\small ( m_i )}} : \tau_i^{( m_i )} )$--store defined
  precisely on $\{ x_i^{{\small ( 1 )}}, \ldots, x_i^{{\small ( m_i )}} \}$.
  Each $\sigma_i$ is thus minimal with this property with respect to the
  standard order on partial functions. Hence

  \[ | P_1, \ldots, P_k | \equallim^{\tmop{def} .} \left( \sigma_i ( x^{p_i}
     )_{_{1 \leqslant i \leqslant k, 1 \leqslant p_i \leqslant n_i}}  \right)
     \in \prod_{_{1 \leqslant i \leqslant k, 1 \leqslant p_i \leqslant n_i}}
     \left\lceil \tau_i^{p_i} \right\rceil \cong \prod_{1 \leqslant j
     \leqslant n} \left\lceil \tau_j \right\rceil \]
  
\end{theproof}

\begin{definition}
  The map $|_{_-} |$ of proposition \ref{prop-storemap} is called the
  {\tmem{store-map}}. The map
  \[ \begin{array}{llll}
       \|_- \| : & \Psi_t & \longrightarrow & \prod_{1 \leqslant i
       \leqslant n} L_{\Gamma}\\
       &  &  & \\
       & \left[ \begin{array}{l}
         P_1\\
         \vdots\\
         P_n
       \end{array} \right] & \longmapsto & \left[ \begin{array}{l}
         l_1\\
         \vdots\\
         l_n
       \end{array} \right]
     \end{array} \]
  is called the {\tmem{location-map}}.
\end{definition}

\section{\label{sec-evo}\!{\tmname{CIP}} Evolutions}

 In this section, we construct HDA's from CIP-programs applying the
operational rules. We then explain how categorification of morphisms of such
HDA's gives rise to ulf-functors. This leads to a finite presentation in terms
of pseudofunctors into $\tmmathbf{\tmop{Span}}$. 

\subsection{From CIP Programs to HDA's}

\subsubsection{Alphabets}

Let
\[ \begin{array}{lll}
     S & \equallim^{\tmop{def} .} & \{ \alpha, \gamma_1, \gamma_2 \}\\
     O & \equallim^{\tmop{def} .} & \{ !_{w, p} | \Gamma \in
     \tmmathbf{\mathcal{K}}, x \in \tmop{Var} ( \Gamma ), w \in \left\lceil
     \Gamma ( x ) \right\rceil, p \in \mathcal{P} \}\\
     I & \equallim^{\tmop{def} .} & \{ ?_{w, p} | \Gamma \in
     \tmmathbf{\mathcal{K}}, x \in \tmop{Var} ( \Gamma ), w \in \left\lceil
     \Gamma ( x ) \right\rceil, p \in \mathcal{P} \}
   \end{array} \]
and
\[ \mathcal{E} \equallim^{\tmop{def} .}  S \cup O \cup I  \]
{\noindent}Given $\mathcal{\hat{E}} \equallim^{\tmop{def} .} \{ \alpha,
\gamma_1, \gamma_2, !, ? \}$, there is the obvious  ``erasure'' map
\[ \begin{array}{lllll}
     \left. \right\|_- \| : & \mathcal{E} & \longrightarrow &
     \widehat{\mathcal{E}} & \\
     & \alpha & \longmapsto & \alpha & \\
     & \gamma_1 & \longmapsto & \gamma_1 & \\
     & \gamma_2 & \longmapsto & \gamma_2 & \\
     & !_{w, p} & \longmapsto & ! & \text{for all} \; w \in \left\lceil
     \Gamma ( x ) \right\rceil\; \text{and}\; p \in \mathcal{P}\\
     & ?_{w, p} & \longmapsto & ? & \text{for all} \; w \in \left\lceil
     \Gamma ( x ) \right\rceil \;\text{and}\; p \in \mathcal{P}
   \end{array} \]

\subsubsection{\label{sec-3-hda}HDA's}

Let $ \Gamma \in \tmmathbf{\mathcal{K}}$ be a
 context, $t \in \mathcal{T}_{\Gamma}$  be a
{\tmname{CIP}} term and 
\[ 
\left( \left[ \text{{\tmem{s}}}_1, \ldots,
\text{{\tmem{s}}}_n ] \right., [ \text{\texttt{p}}_1 \asymp
\text{\texttt{q}}_1, \ldots, \text{\texttt{}} \text{\texttt{}}
\text{\texttt{q}}_m \asymp \text{\texttt{q}}_m ] ) \right.
\] 
be its normal form
(c.f. \ref{sec-3-equterm}). The HDA's
\[ \left\llbracket t \right\rrbracket^{\Gamma} : \left\llbracket t
   \right\rrbracket^{\Gamma, c} \longrightarrow ! \mathcal{E} \]
respectively
\[  \widehat{\left\llbracket t \right\rrbracket}^{\Gamma} :
   \widehat{\left\llbracket t \right\rrbracket}^{\Gamma, c} \longrightarrow !
   \widehat{\mathcal{E}} \]
are constructed as follows.
\begin{enumeratenumeric}
  \item The sets of states are
  \[ \left\llbracket t \right\rrbracket^{\Gamma, c}_0 \;\equallim^{\tmop{def} .}\;
     \{ |P_1, \ldots, P_n | \;{\large |}\; [ P_1 \ldots, P_n ] \in
     \Psi_t \} \]
  respectively
  \[ \widehat{\left\llbracket t \right\rrbracket}^{\Gamma, c}_0
     \;\equallim^{\tmop{def} .}\; \{ \| P_1, \ldots, P_n \| 
\;{\large |}\; [ P_1
     \ldots, P_n ] \in \Psi_t \} \]
  \item  The 1-skeletons are given by the applications of the rewriting rules.
  Each state
  \[ |P_1, \ldots, P_n | \in \left\llbracket t \right\rrbracket^{\Gamma, c}_0
  \]
  is a vertex of a graph $\left\llbracket t \right\rrbracket^{\Gamma, c}_{0,
  1} \;\equallim^{\tmop{def} .}\;  \left( \left\llbracket t
  \right\rrbracket^{\Gamma, c}_0, \left\llbracket t \right\rrbracket^{\Gamma,
  c}_1 \right)$. Each edge of this graph comes from an application of some
  rewrite rule
  \[ \frac{\varpi}{[ P_1, \ldots, P_n ] \longrightarrow [ P_1', \ldots, P_n'
     ]} \]
  so it can be labeled according to the latter. This labeling, written
  \[ \left\llbracket t \right\rrbracket^{\Gamma}_{0, 1} : \left\llbracket t
     \right\rrbracket^{\Gamma, c}_{0, 1} \longrightarrow \mathcal{E} \]
  is given by
  \[ \begin{array}{|l|l|l|l|l|l|}
       \hline
       \text{{\tmstrong{{\tmname{Asg}}}}} & \text{{\tmstrong{{\tmname{If1}}}}}
       / \text{{\tmstrong{{\tmname{While1}}}}} &
       \text{{\tmstrong{{\tmname{If2}}}}} /
       \text{{\tmstrong{{\tmname{While2}}}}} & \text{{\tmstrong{RV}}} &
       \text{{\tmstrong{S}}} & \text{{\tmstrong{R}}} \hspace{0.25em}_w\\
       \hline
       \alpha & \gamma_1 & \gamma_2 & \alpha & !_{\sigma_i ( e ), \mathtt{p}}
       & ?_{w, \mathtt{q}}\\
       \hline
     \end{array} \]
  Similarly, the labeling of the graph $\widehat{\left\llbracket t
  \right\rrbracket}^{\Gamma, c}_{0, 1}$, written
  \[ \widehat{\left\llbracket t \right\rrbracket}^{\Gamma}_{0, 1} :
     \widehat{\left\llbracket t \right\rrbracket}^{\Gamma, c}_{0, 1}
     \longrightarrow \widehat{\mathcal{E}} \]
  is given by
  \[ \begin{array}{|l|l|l|l|l|l|}
       \hline
       \text{{\tmstrong{{\tmname{Asg}}}}} & \text{{\tmstrong{{\tmname{If1}}}}}
       / \text{{\tmstrong{{\tmname{While1}}}}} &
       \text{{\tmstrong{{\tmname{If2}}}}} /
       \text{{\tmstrong{{\tmname{While2}}}}} & \text{{\tmstrong{RV}}} &
       \text{{\tmstrong{S}}} & \text{{\tmstrong{R}}} \hspace{0.25em}_w\\
       \hline
       \alpha & \gamma_1 & \gamma_2 & \alpha & !_{} & ?\\
       \hline
     \end{array} \]
  Degeneracies are added freely and labeled with $\star$
  
  \item $\left\llbracket t \right\rrbracket^{\Gamma}$ is  the 1-coskeletal HDA
  \[ \left\llbracket t \right\rrbracket^{\Gamma} \;\equallim^{\tmop{def} .}\;  
     \tmop{cosk}_{\mathcal{E}} \left( \left\llbracket t
     \right\rrbracket^{\Gamma}_{0, 1} \right) \]
  while $\widehat{\left\llbracket t \right\rrbracket}^{\Gamma}$ is  the
  1-coskeletal HDA
  \[  \widehat{\left\llbracket t \right\rrbracket}^{\Gamma}
     \;\equallim^{\tmop{def} .}\;   \tmop{cosk}_{\mathcal{\hat{E}}} \left(
     \widehat{\left\llbracket t \right\rrbracket}^{\Gamma}_{0, 1} \right) \]
\end{enumeratenumeric}
\begin{proposition}
  \label{prop-stacub} $\left\llbracket s \right\rrbracket^{\Gamma, c}_{0, 1}$
  and  $\widehat{\left\llbracket s \right\rrbracket} ^{\Gamma, c}_{0, 1}$ are
  acubic for all $s \in \mathcal{L}_{\Gamma}$.
\end{proposition}

\begin{theproof}
  Induction on syntax.
\end{theproof}

\begin{proposition}
  \label{prop-stats}The 1-skeletal HDA's $\left\llbracket s
  \right\rrbracket^{\Gamma}$ and $\widehat{\left\llbracket s
  \right\rrbracket}^{\Gamma}$  are transition systems, i.e.
  \[ \forall e, f \in \left\llbracket s \right\rrbracket_1^{\Gamma, c} .
        \left\llbracket s \right\rrbracket_1^{\Gamma} ( e ) = \left\llbracket
        s \right\rrbracket_1^{\Gamma} ( f ) \Rightarrow e = f \] 
  respectively
  \[ \forall e, f \in \widehat{\left\llbracket s \right\rrbracket}_1^{\Gamma,
     c} . \widehat{\left\llbracket s \right\rrbracket_1}^{\Gamma} ( e ) =
     \widehat{\left\llbracket s \right\rrbracket_1}^{\Gamma} ( f ) \Rightarrow
     e = f \]
  for all $s \in \mathcal{L}_{\Gamma}$.
\end{proposition}

\begin{theproof}
  Induction on syntax.
\end{theproof}

\begin{definition}
  \label{def-synctab} Let $t \in \mathcal{T}_{_{\Gamma}}$ be a
  term with normal form 
\[ \left( 
[s_1,\ldots,s_n],[\mathtt{p}_1 \asymp \mathtt{q}_1,\ldots,
\mathtt{p}_1 \asymp \mathtt{q}_1]
\right)\]
The term $t$ determines the synchronization table 
$\mathbf{s}_t (_-,_- )$ such that
  \begin{enumerateroman}
    \item $\tmop{Tab}_t ( \star, \theta ) = \tmop{Tab}_t ( \theta, \star )
    \;\equallim^{\tmop{def} .}\; \theta$ for all $\theta \in \mathcal{E}$;
    
    \item $\tmop{Tab}_t ( !_{v, \mathtt{p} )}, ?_{( w, \mathtt{q} )} ) =
    \tmop{Tab}_t ( ?_{( w, \mathtt{q} )}, !_{v, \mathtt{p} )} )
    \;\equallim^{\tmop{def} .}\; \alpha$ if $\mathtt{p} \asymp \mathtt{q}$ and $v= w$;
    
    \item all other entries are $\top$.
  \end{enumerateroman}
\end{definition}

We write $\boxast_t$ for the synchronized product of HDA's with respect to
$\mathbf{s}_t$ and $\boxtimes_t$ for its counterpart on the underlying
reflexive graphs.

\begin{proposition}
  Suppose $t \in \mathcal{T}_{_{\Gamma}}$  with normal form
\[ \left( 
[s_1,\ldots,s_n],[\mathtt{p}_1 \asymp \mathtt{q}_1,\ldots,
\mathtt{p}_1 \asymp \mathtt{q}_1]
\right)\]
and let $\Gamma_i \in
  \tmmathbf{\mathcal{K}}$ be the contexts such that $s_i \in
  \mathcal{T}_{_{\Gamma_i}}$  for all $1 \leqslant i \leqslant n$. Then
  \[ \left\llbracket t \right\rrbracket^{\Gamma} \cong \left\llbracket s_1
        \right\rrbracket^{\Gamma_1} \boxast_t \cdots \boxast_t \left\llbracket
        s_n \right\rrbracket^{\Gamma_n} \] 
  respectively
  \[ \widehat{\left\llbracket t \right\rrbracket}^{\Gamma} \cong
     \widehat{\left\llbracket s_1 \right\rrbracket}^{\Gamma_1} \boxast_t
     \cdots \boxast_t \widehat{\left\llbracket s_n
     \right\rrbracket}^{\Gamma_n} \]
\end{proposition}

\begin{theproof}
  Analyzing the rewrite rules, it is easy to see that
   \[ \left\llbracket t \right\rrbracket^{\Gamma, c}_{0, 1} \cong
        \left\llbracket s_1 \right\rrbracket^{\Gamma_1, c}_{0, 1} \boxtimes_t
        \cdots \boxtimes_t \left\llbracket s_n \right\rrbracket^{\Gamma_n,
        c}_{0, 1} \] 
  respectively
  \[ \widehat{\left\llbracket t \right\rrbracket}^{\Gamma, c}_{0, 1} \cong
     \widehat{\left\llbracket s_1 \right\rrbracket}^{\Gamma_1, c}_{0, 1}
     \boxtimes_t \cdots \boxtimes_t \widehat{\left\llbracket s_n
     \right\rrbracket}^{\Gamma_n, c}_{0, 1} \]
  
\end{theproof}

\begin{proposition}
  \label{prop-pimor}There is a morphism of reflexive graphs
  \[ \begin{array}{llll}
       \pi_s : & \left\llbracket s \right\rrbracket^{\Gamma, c}_{0, 1} &
       \longrightarrow & \widehat{\left\llbracket s \right\rrbracket}^{\Gamma,
       c}_{0, 1}\\
       &  &  & \\
       & |P| & \longmapsto & \left. \| P \right\|\\
       & |P| \rightarrowlim^{\kappa} |P' | & \longmapsto & \| P \|
       \longrightarrowlim^{\| \kappa \|} \| P' \|
     \end{array} \]
for all $s \in \mathcal{L}_{\Gamma}$.
\end{proposition}

\begin{theproof}
  The morphism $\pi_s$ is well-defines on edges by proposition
  \ref{prop-stats}.
\end{theproof}

\begin{remark}
  \label{rem-cipcommute}The diagram
  
  \begin{center}
    $
\xymatrix @*[r]{[\![s]\!]^{\Gamma,c}_{0,1} \ar[r]^{\pi_s} 
\ar[d]_{[\![s]\!]^{\Gamma}_{0,1}}& 
\widehat{[\![s]\!]}^{\Gamma,c}_{0,1} 
\ar[d]^{\widehat{[\![s]\!]}^{\Gamma}_{0,1}} \\
\mathcal{E} \ar[r]_{\|\_\|} & \hat{\mathcal{E}}
}
$

  \end{center}
  
  {\noindent}commutes by construction.
\end{remark}

\subsection{Categorification as a generalized Relational Semantics}
\label{sec-2-categorific}

\begin{theorem}
  \label{theo-ciprelative}Suppose $t \in \mathcal{T}_{_{\Gamma}}$  with normal
  form  $( [ \text{{\tmem{s}}}_1, \ldots, \text{{\tmem{s}}}_n ], [
  \text{\texttt{p}}_1 \asymp \text{\texttt{q}}_1, \ldots, \text{\texttt{}}
  \text{\texttt{}} \text{\texttt{q}}_m \asymp \text{\texttt{q}}_m ] )$. Let
  $\Gamma_i \in \tmmathbf{\mathcal{K}}$ be the contexts such that $s_i \in
  \mathcal{T}_{_{\Gamma_i}}$  for all $1 \leqslant i \leqslant n$ and
  $\pi_{s_i}$ be the morphism of proposition \ref{prop-pimor} for all $1
  \leqslant i \leqslant n$. Let
  \[ \bar{w}_{i, i + 1} : \mathcal{E}_{\left\llbracket s_i
     \right\rrbracket^{\Gamma_i}_{0, 1}, \left\llbracket s_{i + 1}
     \right\rrbracket^{\Gamma_{i + 1}}_{0, 1}} \longrightarrow
     \mathcal{\hat{E}}_{\left\llbracket s_i \right\rrbracket^{\Gamma_i}_{0,
     1}, \left\llbracket s_{i + 1} \right\rrbracket^{\Gamma_{i + 1}}_{0, 1}}
  \]
  be as in theorem \ref{theo-relative}. Then
  \[ ( \mathcal{C} \circ \tmop{cosk}_{\Sigma} ) ( \pi_{s_1} \boxtimes_t \cdots
     \boxtimes_t \pi_{s_n} ) \cong \mathcal{F} ( \pi_{s_1} )
     \times_{\mathcal{}_{} \mathcal{F} ( \bar{w}_{1, 2} )}  \mathcal{F} (
     \pi_{s_2} ) \times_{_{\mathcal{F} ( \bar{w}_{2, 3} )}} \cdots
     \times_{_{\mathcal{F} ( \bar{w}_{n - 1, n} )}} \mathcal{F} ( \pi_{s_n} )
  \]
\end{theorem}

\begin{theproof}
  By proposition \ref{prop-stacub}, remark \ref{rem-cipcommute} and theorem
  \ref{theo-relative}.
\end{theproof}

Theorem \ref{theo-ciprelative} produces a workable abstraction of the more
precise HDA-semantics, the degree of precision being measured by the unit of
the adjunction $\mathcal{C} \dashv \mathcal{N}$ (c.f. proposition
\ref{prop-cubnerve}). This abstraction is the limit $\pi_t : E_t
\longrightarrow C_t$ of the diagram

$
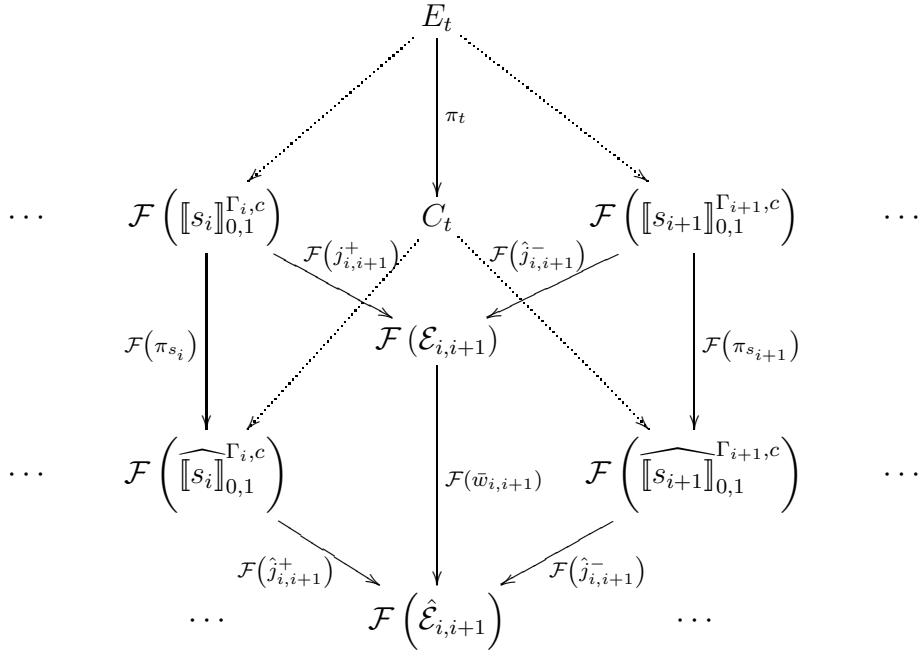
\xymatrix { && E_t \ar[dd]^{\pi_t} \ar@{.>}[ddl] \ar@{.>}[ddr]&& \\
&&&&
\\
\cdots &  \mathcal{F}\left([\![ s_i ]\!]^{\Gamma_i,c}_{0,1}\right) 
\ar[dd]_{\mathcal{F}\left(\pi_{s_i}\right)} 
\ar[dr]^{\free\left(j^+_{i,i+1}\right)} 
&
C_t \ar@{.>}[ddl] \ar@{.>}[ddr] 
&
\mathcal{F}\left([\![ s_{i+1} ]\!]^{\Gamma_{i+1},c}_{0,1}\right) 
\ar[dd]^{\mathcal{F}\left(\pi_{s_{i+1}}\right)} 
\ar[dl]_{\free\left(\hat{j}^-_{i,i+1}\right)}
 & \cdots
\\
 &&  \mathcal{F}\left(\mathcal{E}_{i,i+1}\right)  
\ar[dd]^{\free (\bar{w}_{i,i+1})} &&  
\\
\cdots & \mathcal{F}\left(\widehat{[\![ s_i ]\!]}^{\Gamma_i,c}_{0,1}\right)  
\ar[dr]_{\free\left(\hat{j}^+_{i,i+1}\right)} &&
\mathcal{F}\left(\widehat{[\![ s_{i+1} ]\!]}^{\Gamma_{i+1},c}_{0,1}\right)   
\ar[dl]^{\free\left(\hat{j}^-_{i,i+1}\right)} & \cdots
\\
 &\cdots&  \mathcal{F}\left(\hat{\mathcal{E}}_{i,i+1}\right)  &\cdots& 
}
$

{\noindent}in  $\tmmathbf{\tmop{Cat}}^{\rightarrow}$ where
\begin{itemizeminus}
  \item $\mathcal{E}_{i, i + 1} \;\equallim^{\tmop{def} .}\;
  \mathcal{E}_{\left\llbracket s_i \right\rrbracket^{\Gamma_i}_{0, 1},
  \left\llbracket s_{i + 1} \right\rrbracket^{\Gamma_{i + 1}}_{0, 1}}
  \text{ and } \widehat{\mathcal{E}}_{i, i + 1} \;\equallim^{\tmop{def} .}\;
  \widehat{\mathcal{E}}_{\left\llbracket s_i \right\rrbracket^{\Gamma_i}_{0,
  1}, \left\llbracket s_{i + 1} \right\rrbracket^{\Gamma_{i + 1}}_{0, 1}}$
  
  \item $j^+_{i, i + 1} \;\equallim^{\tmop{def} .}\; \left\llbracket s_i
  \right\rrbracket^{\Gamma_i}_{0, 1} \circ j^{\left\llbracket s_i
  \right\rrbracket^{\Gamma_i, c}_{0, 1}}_{\left\llbracket s_i
  \right\rrbracket^{\Gamma_i}_{0, 1}, \left\llbracket s_{i + 1}
  \right\rrbracket^{\Gamma_{i + 1}}_{0, 1}} \text{ and } \hat{j}^+_{i, i + 1}
  \;\equallim^{\tmop{def} .}\; \widehat{\left\llbracket s_i
  \right\rrbracket}^{\Gamma_i}_{0, 1} \circ j^{\widehat{\left\llbracket s_i
  \right\rrbracket}^{\Gamma_i, c}_{0, 1}}_{\widehat{\left\llbracket s_i
  \right\rrbracket}^{\Gamma_i}_{0, 1}, \widehat{\left\llbracket s_{i + 1}
  \right\rrbracket}^{\Gamma_{i + 1}}_{0, 1}} $
  
  \item $j^-_{i, i + 1} \;\equallim^{\tmop{def} .}\; \left\llbracket s_{i + 1}
  \right\rrbracket^{\Gamma_{i + 1}}_{0, 1} \circ j^{\left\llbracket s_{i + 1}
  \right\rrbracket^{\Gamma_{i + 1}, c}_{0, 1}}_{\left\llbracket s_i
  \right\rrbracket^{\Gamma_i}_{0, 1}, \left\llbracket s_{i + 1}
  \right\rrbracket^{\Gamma_{i + 1}}_{0, 1}} \text{ and } \hat{j}^-_{i, i + 1}
  \;\equallim^{\tmop{def} .}\; \widehat{\left\llbracket s_{i + 1}
  \right\rrbracket}^{\Gamma_{i + 1}}_{0, 1} \circ j^{\widehat{\left\llbracket
  s_{i + 1} \right\rrbracket}^{\Gamma_{i + 1}, c}_{0,
  1}}_{\widehat{\left\llbracket s_i \right\rrbracket}^{\Gamma_i}_{0, 1},
  \widehat{\left\llbracket s_{i + 1} \right\rrbracket}^{\Gamma_{i + 1}}_{0,
  1}} $
\end{itemizeminus}
(c.f. theorem \ref{theo-catsynchro}). This diagram is actually in 
$\tmmathbf{\tmop{Ulf}}_{\mathcal{F}}^{\rightarrow}$ by remark
\ref{rem-ulf}. We call
\[ E_t \cong  \mathcal{F} \left( \left\llbracket s_1
   \right\rrbracket^{\Gamma_1}_{0, 1} \right) 
\times_{\mathcal{F}(\mathcal{E}_{1,2})}
\mathcal{F} \left( \left\llbracket s_2
   \right\rrbracket^{\Gamma_2}_{0, 1} \right) \times_{\mathcal{}_{\mathcal{F}
   ( \mathcal{E}_{1, 2} )}} \cdots \times_{\mathcal{}_{\mathcal{F} (
   \mathcal{E}_{n - 1, n} )}} \mathcal{F} \left(
   \left\llbracket s_n \right\rrbracket^{\Gamma_n}_{0, 1} \right) \]
$t$'s {\tmem{category of evolutions}} and
\[ C_t \cong \mathcal{F} \left( \widehat{\left\llbracket s_1
   \right\rrbracket}^{\Gamma_1}_{0, 1} \right) \times_{_{\mathcal{F} (
   \mathcal{\hat{E}}_{1, 2} )}} \mathcal{F} \left(
   \widehat{\left\llbracket s_2 \right\rrbracket}^{\Gamma_2}_{0, 1} \right)
   \times_{\mathcal{}_{_{\mathcal{F} ( \mathcal{\hat{E}}_{1, 2} )}}} \cdots
   \times_{\mathcal{}_{\mathcal{F} ( \mathcal{\hat{E}}_{n - 1, n}
   )}} \mathcal{F} \left( \widehat{\left\llbracket s_n
   \right\rrbracket}^{\Gamma_n}_{0, 1} \right) \]
$t$'s {\tmem{control category}}. Accordingly, we call $E_i
\;\equallim^{\tmop{def} .}\;  \mathcal{F} \left( \left\llbracket s_i
\right\rrbracket^{\Gamma_i}_{0, 1} \right)$ and $C_i \;\equallim^{\tmop{def} .}\; 
\mathcal{F} \left( \widehat{\left\llbracket s_i
\right\rrbracket}^{\Gamma_i}_{0, 1} \right)$ $s_i$'s category of evolutions
respectively $s_i$'s control category. $I_{i, i + 1} \;\equallim^{\tmop{def} .}\; 
\mathcal{F} \left( \mathcal{\hat{E}}_{i, i + 1} \right)$ is the control
category of the corresponding {\tmem{interface}} $\phi_{i, i + 1}
\;\equallim^{\tmop{def} .}\;  \mathcal{F} ( \bar{w}_{i, i + 1} )$.

\begin{proposition}
  \label{prop-finite}The control category $C_i$ is  finite  for all 
$1 \leqslant i \leqslant n$. In particular, the control category $C_t$ is 
finite.
\end{proposition}

\begin{theproof}
  Induction on syntax.
\end{theproof}

Let us abuse notation by writing $\pi_{s_i}$ for $\mathcal{F} ( \pi_{s_i} )$.
The Giraud-Conduch\'e correspondence 
(c.f. section \ref{sec-2-thecorr}) produces
a finite presentation of $\pi_t$ as a pseudofunctor 
\[\underline{\pi_t} : C_t
\longrightarrow \tmmathbf{\tmop{Span}}\]
This pseudofunctor is a wide pullback
object
\[ \underline{\pi_t} \cong \underline{\pi_{s_1}} \times_{\underline{\phi_{1,
   2}}}  \underline{\pi_{s_2}} \times_{\underline{\phi_{2, 3}}} \cdots
   \times_{\underline{\phi_{n - 1, n}}}  \underline{\pi_{s_n}}  \]
in $\mathcal{F} / \! /\mathbf{Span}$. On the other hand,
the hypothesis on $t$ and proposition \ref{prop-nform} entail that
\[ \text{$t \approx s_1 \ll \Phi_{1, 2} \gg s_2 \ll \Phi_{2, 3} \gg \cdots \ll
   \Phi_{n - 1, n} \gg s_n$} \]
for lists of channels $\Phi_{i, i + 1}$. Suppose $\Gamma_i =
\text{\texttt{x}}_1 : \tau_1, \ldots, \text{\texttt{x}}_{m_i} : \tau_{m_i}$
and let
\[ \left\lceil \Gamma_i \right\rceil \;\equallim^{\tmop{def} .}\;  \prod_{1
   \leqslant j \leqslant m_i}  \left\lceil \tau_j \right\rceil \]
for all $1 \leqslant i \leqslant n$. Unraveling theorem \ref{theo-gircond} and
theorem \ref{theo-ciprelative} gives the following characterization of
$\underline{\pi_t}$.
\begin{enumerate}
  \item There is \[\underline{\phi_{i, i + 1}} : I_{i, i + 1} \longrightarrow
  \tmmathbf{\tmop{Span}}\] for all $1 \leqslant i \leqslant n$. $I_{i, i + 1}$
  has one vertex and is generated by $\Phi_{i, i + 1}$. Each generator
  $\text{\texttt{p}} \asymp \text{\texttt{q}}_{} \in \Phi_{i, i + 1} $has an
  associated type $\tau$ since it comes from an application of
  $\text{{\tmstrong{MCut{\tmname{}}}}}$. We have
  \[ \underline{\phi_{i, i + 1}} ( \text{\texttt{p}} \asymp \text{\texttt{q}}
     ) = ( !, \left\lceil \tau \right\rceil, ! ) \]
  where $( !, \left\lceil \tau \right\rceil, ! )$ is the span
  \[ 1 \longleftarrowlim^! \left\lceil \tau \right\rceil \longrightarrowlim^!
     1 \]
  in $\tmmathbf{\tmop{Set}}$. We have further $\underline{\phi_{i, i + 1}} ( \tmop{id} ) =
  ( !, \left\lceil \Gamma_i \right\rceil, ! )$.
  
  \item There is \[\underline{\pi_{s_i}} : C_{s_i} \longrightarrow
  \tmmathbf{\tmop{Span}}\] for all $1 \leqslant i \leqslant n$. $C_{s_i}$'s
  vertices are program locations. Then
  \[ \left( \underline{\pi_{s_i}} \right)_0 ( l ) \equallim \left\lceil
     \Gamma_i \right\rceil \]
  for all $1 \leqslant i \leqslant n$. Let $l \in \left( C_{s_i} \right)_0$
  and $\psi ( l )$ be the statement at location $l$. Given the syntax and the
  operational rules, there are the following possible cases.
  \begin{enumerate}
    \item $\text{out} ( l ) = \varnothing$;
    
    \item \label{B}if $\text{out} ( l ) = \{ l_1 \}$ then there are four
    cases;
    \begin{enumerate}
      \item if $\psi ( l ) \equiv \text{}^{\backprime \backprime}
      \mathtt{\tmop{nop}} \text{}^{\backprime \backprime}$ then
      \[ \underline{\pi_{s_i}} ( l, l_1 ) = ( \tmop{id}, \left\lceil \Gamma_i
         \right\rceil, \tmop{id} ) ; \]
      \item if $\psi ( l ) \equiv \text{}^{\backprime \backprime} \mathtt{x_j}
      : = e \text{}^{\backprime \backprime}$ then
      \[ \underline{\pi_{s_i}} ( l, l_1 ) = ( \tmop{id}, \left\lceil \Gamma_i
         \right\rceil, \pi_1 \times \cdots \times \pi_{j - 1} \times
         \left\lceil e \right\rceil \times \pi_{j + 1} \times \cdots \times
         \pi_{m_i} ; \]
      \item \label{Bc}if $\psi ( l ) \equiv \text{}^{\backprime \backprime}
      \mathtt{p} !e \text{}^{\backprime \backprime}$ then
      \[ \underline{\pi_{s_i}} ( l, l_1 ) = ( \tmop{id}, \left\lceil \Gamma_i
         \right\rceil, \tmop{id} ) ; \]
      \item if $\psi ( l ) \equiv \text{}^{\backprime \backprime} \mathtt{p} ?
      \mathtt{x} \text{}^{\backprime \backprime}$ then
      \[ \underline{\pi_{s_i}} ( l, l_1 ) = ( \pi_1, \left\lceil \Gamma_i
         \right\rceil \times \left\lceil \Gamma_i \right\rceil, \pi_2 ) ; \]
    \end{enumerate}
    \item if $\text{out} ( l ) = \{ l_1, l_2 \}$ then there are two cases;
    \begin{enumerate}
      \item if $\psi ( l ) \equiv \text{}^{\backprime \backprime}
      \mathtt{\tmop{if}} \hspace{0.5em} e \hspace{0.5em} \mathtt{\tmop{then}}
      \hspace{0.5em} \text{{\tmem{stat}}}_1 \hspace{0.5em}
      \mathtt{\tmop{else}} \hspace{0.5em} \text{{\tmem{stat}}}_2
      \hspace{0.5em} \mathtt{\tmop{end}} \text{}^{\backprime \backprime}$ ;
      
      \item $\mathtt{\psi ( l ) \equiv \text{}^{\backprime \backprime}
      \tmop{while}}  \hspace{0.5em} e \hspace{0.5em} \mathtt{\tmop{do}}
      \hspace{0.5em} \text{{\tmem{stat}}} \hspace{0.5em} \mathtt{\tmop{end}}
      \text{}^{\backprime \backprime}$ ;
    \end{enumerate}
    in both cases, suppose w.l.o.g that $\widehat{\left\llbracket t
    \right\rrbracket}^{\Gamma}_{0, 1} ( l, l_1 ) = \gamma_1$ and
    $\widehat{\left\llbracket t \right\rrbracket}^{\Gamma}_{0, 1} ( l, l_2 ) =
    \gamma_2$ (c.f. section \ref{sec-3-hda}). Let $\kappa : S_e
    \rightarrowtail \left\lceil \Gamma_i \right\rceil$ be the subobject
    classified by $\left\lceil e \right\rceil : \left\lceil \Gamma_i
    \right\rceil \longrightarrow \{ \tmop{tt}, \tmop{ff} \},$ that is
    \[ S_e \;\equallim^{\tmop{def} .}\; \{ x \in \left\lceil \Gamma_i \right\rceil
       | \left\lceil e \right\rceil ( x ) \} \]
    Let $\bar{\kappa} : \overline{S_e} \rightarrowtail \left\lceil \Gamma_i
    \right\rceil$ be its complement. Then
    \[ \underline{\pi_{s_i}} ( l, l_1 ) = ( \kappa, S_e, \kappa ) \]
    and
    \[ \underline{\pi_{s_i}} ( l, l_2 ) = ( \bar{\kappa}, \overline{S_e},
       \bar{\kappa} ) . \]
    
  \end{enumerate}
  \item There are lax representable transformations
  \[ \iota_{i, i + 1}^- \;\equallim^{\tmop{def} .}\;  \underline{\left(
     \mathcal{F} ( j_{i, i + 1}^- ), \mathcal{F} ( \hat{j}_{i, i + 1}^- )
     \right)} : \underline{\pi_{s_{i + 1}}} \Longrightarrow
     \underline{\phi_{i, i + 1}} \circ \mathcal{F} ( \hat{j}_{i, i + 1}^- )
  \]
  and
  \[ \iota_{i, i + 1}^+ \;\equallim^{\tmop{def} .}\;  \underline{\left(
     \mathcal{F} ( j_{i, i + 1}^+ ), \mathcal{F} ( \hat{j}_{i, i + 1}^+ )
     \right) :}  \underline{\pi_{s_i}} \Longrightarrow \underline{\phi_{i, i +
     1}} \circ \mathcal{F} ( \hat{j}_{i, i + 1}^+ ) \]
  for all $1 \leqslant i < n$. Let
  \[ \Gamma_{i, i + 1}^{\omega} \;\equallim^{\tmop{def} .}\; 
     \left\{\begin{array}{l}
       \begin{array}{ll}
         \Gamma_{i + 1} & \omega = -\\
         \Gamma_i & \omega = +
       \end{array}
     \end{array}\right. \]
  We have
  \[ ( \iota_{i, i + 1}^{\omega} )_l = ( \tmop{id},\left\lceil
     \Gamma^{\omega}_{i, i + 1} \right\rceil, ! ) \]
  for all $l \in \left( C_{s_i} \right)_0$ , $\omega \in \{ -, + \}$ and $1
  \leqslant i < n$. Given a generator $( l, l' ) \in C_{s_i}$, there are five
  cases:
  \begin{enumerate}
    \item if $\underline{\pi_{s_i}} ( l, l' ) = ( \tmop{id}, \left\lceil
    \Gamma^{\omega}_{i, i + 1} \right\rceil, \tmop{id} )$ and $\psi ( l )
    \equiv \text{}^{\backprime \backprime} \mathtt{\tmop{nop}}
    \text{}^{\backprime \backprime}$ then
    \[ ( \iota_{i, i + 1}^{\omega} )_{( l, l' )} = \tmop{id}_{\left\lceil
       \Gamma^{\omega}_{i, i + 1} \right\rceil} : ( \tmop{id}, \left\lceil
       \Gamma^{\omega}_{i, i + 1} \right\rceil, \tmop{id} ) \longrightarrow (
       \tmop{id}, \left\lceil \Gamma^{\omega}_{i, i + 1} \right\rceil,
       \tmop{id} ) ; \]
    this morphism of spans corresponds to the case 2.(b).i above;
    
    \item if $\underline{\pi_{s_i}} ( l, l' ) = ( \tmop{id}, \left\lceil
    \Gamma^{\omega}_{i, i + 1} \right\rceil, \pi_1 \times \cdots \times \pi_{j
    - 1} \times \left\lceil e \right\rceil \times \pi_{j + 1} \times \cdots
    \times \pi_{m_{}} )$ where
    \[ m = \left\{\begin{array}{ll}
         m_{i + 1} & \omega = -\\
         m_i & \omega = +
       \end{array}\right. \]
    then
    \[ ( \iota_{i, i + 1}^{\omega} )_{( l, l' )} = \tmop{id}_{\left\lceil
       \Gamma^{\omega}_{i, i + 1} \right\rceil} : ( \tmop{id}, \left\lceil
       \Gamma^{\omega}_{i, i + 1} \right\rceil, ! ) \longrightarrow (
       \tmop{id}, \left\lceil \Gamma^{\omega}_{i, i + 1} \right\rceil, ! ) ;
    \]
    this morphism of spans corresponds to the case 2.(b).ii above;
    
    \item if $\underline{\pi_{s_i}} ( l, l' ) = ( s, S, s )$ for a subobject
    $s : S \rightarrowtail \left\lceil \Gamma^{\omega}_{i, i + 1}
    \right\rceil$then
    \[ ( \iota_{i, i + 1}^{\omega} )_{( l, l' )} = s : ( s, S, ! )
       \longrightarrow ( \tmop{id}, \left\lceil \Gamma^{\omega}_{i, i + 1}
       \right\rceil, ! ) ; \]
    this morphism of spans corresponds to cases  2.(c).i and 2.(c).ii above;
    
    \item if $\underline{\pi_{s_i}} ( l, l' ) = ( \tmop{id}, \left\lceil
    \Gamma^{\omega}_{i, i + 1} \right\rceil, \tmop{id} )$ and $\psi ( l )
    \equiv \text{}^{\backprime \backprime} \mathtt{p} !e \text{}^{\backprime
    \backprime}$ then
    \[ ( \iota_{i, i + 1}^{\omega} )_{( l, l' )} = \langle \tmop{id},
       \left\lceil e \right\rceil \rangle : ( \tmop{id}, \left\lceil
       \Gamma^{\omega}_{i, i + 1} \right\rceil, ! ) \longrightarrow ( \pi_1,
       \left\lceil \Gamma^{\omega}_{i, i + 1} \right\rceil \times \left\lceil
       \tau \right\rceil, ! ) ; \]
    this morphism of spans corresponds to the case 2.(a).iii above;
    
    \item if $\underline{\pi_{s_i}} ( l, l' ) = ( \pi_1, \left\lceil
    \Gamma^{\omega}_{i, i + 1} \right\rceil \times \left\lceil
    \Gamma^{\omega}_{i, i + 1} \right\rceil, \pi_2 )$ then
    \[ ( \iota_{i, i + 1}^{\omega} )_{( l, l' )} = \tmop{id} \times \pi_j : (
       \pi_1, \left\lceil \Gamma^{\omega}_{i, i + 1} \right\rceil \times
       \left\lceil \Gamma^{\omega}_{i, i + 1} \right\rceil, ! )
       \longrightarrow ( \pi_1, \left\lceil \Gamma^{\omega}_{i, i + 1}
       \right\rceil \times \left\lceil \tau \right\rceil, ! ) ; \]
    this morphism of spans corresponds to the case 2.(b).iii
 above so $\psi ( l )
    \equiv \text{}^{\backprime \backprime} \mathtt{p} ? \mathtt{x}_j
    \text{}^{\backprime \backprime}$ and $\pi_j : \left\lceil
    \Gamma^{\omega}_{i, i + 1} \right\rceil \longrightarrow \left\lceil \tau_j
    \right\rceil $.
  \end{enumerate}
  
\end{enumerate}
In short, we label the control in the pseudo-monoid $\tmmathbf{\tmop{Span}} (
\left\lceil \Gamma_i \right\rceil, \left\lceil \Gamma_i \right\rceil )$. The
span labeling a control transition encodes all the evolutions ``above'' the
latter. The lax natural transformations to the interface are functional
simulations encoding the observability of the control transitions. For this
reason, objects of the lax comma category 
$\mathcal{F} / \! / \tmmathbf{\tmop{Span}}$ 
are also called {\tmem{categorical transition
systems}} {\cite{linds,lindspap,me,kw:ctcs,kw:cmcim,koslo}}.

\begin{remark}
  \label{rem-gluing} $C_{s_i}$ can be obtained directly from the syntax, by a
  series of pushouts starting from the base cases. The above characterization
  can thus be taken as the definition of a generalized relational semantics of
  {\tmname{CIP}}. The Giraud-Conduch\'e correspondence guarantees ``soundness
  and completeness'' with respect to the semantics formulated in terms of
  categories of evolutions hence ''soundess'' with respect to the semantics
  based on HDA'a.
\end{remark}

{\noindent}For example, the purely sequential {\tmname{CIP}}-program

\texttt{x:$\tmop{nat} \vdash$x:=20; while x>0 do x:= x-1 end$: \langle
\rangle$}

{\noindent}gives rise to the categorical transition system

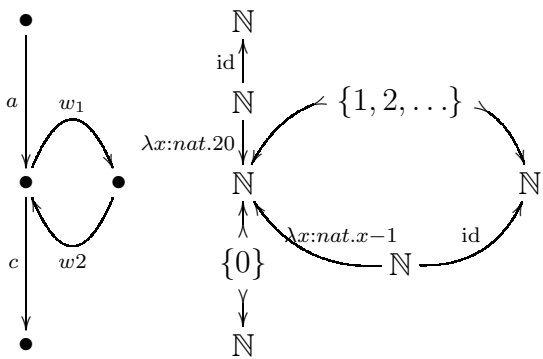
\begin{center}
  $\xymatrix@=3mm{
\bullet \ar[dd]_{a} &  &&& \N &&&
\\
&&&& \N \ar[u]^{\one} \ar[d]_{\lambda x:nat.20} &
**{++} \{1,2,\ldots\}
\ar@/_1.2pc/@{>->}[dl] \ar@/^1.2pc/@{>->}[dr] &&
\\
\bullet \ar[dd]_{c} \ar@/^2pc/[rr]^{w_1} & &
\bullet \ar@/^2pc/[ll]^{w2} & & \N & & \N &
\\
&&&& **{++} \{0\} \ar@{>->}[u] \ar@{>->}[d] & 
\N \ar@/^1pc/[ul]_{\lambda x:nat.x-1} \ar@/_1pc/[ur]^{\one} &&
\\
\bullet & && & \N &&&
}$
\end{center}

{\noindent}while the parallel composition

\texttt{x:$\tmop{nat}$,z:$\tmop{nat} \vdash$x:=5; p!(x+x) $\ll$p$\asymp$q$\gg$
q?z; z:=z*z: $\langle \rangle$}

{\noindent}gives rise to the limit of the diagram

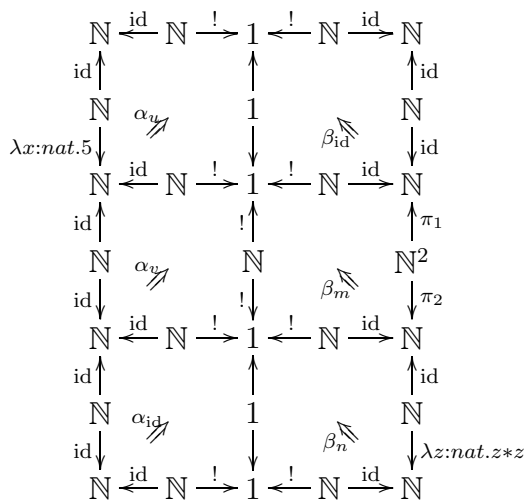
\begin{center}
  $
\xymatrix @=5mm{
        \N & \N \ar[l]_{\one} \ar[r]^{!} \xtwocell[1,1]{}\omit{^<6>{\alpha_u}}
	& \onee & \N \ar[l]_{!} \ar[r]^{\one}
        \xtwocell[1,-1]{}\omit{<-6>{\beta_{\one}}}& \N
	\\
	\N \ar[u]^{\one} \ar[d]_{\lambda x:nat.5}	& 
	&\onee \ar[u] \ar[d]& & \N \ar[u]_{\one} \ar[d]^{\one}
	\\
	\N & 
        \N \ar[l]_{\one} \ar[r]^{!} 
        \xtwocell[1,1]{}\omit{^<6>{\alpha_v}} & 
        \onee &
        \N \ar[l]_{!} \ar[r]^{\one}
        \xtwocell[1,-1]{}\omit{<-6>{\beta_m}} &
        \N
	\\
        \N \ar[u]^{\one} \ar[d]_{\one} & & 
	\N \ar[u]^{!} \ar[d]_{!} & &
        \N^2 \ar[u]_{\pi_1} \ar[d]^{\pi_2}
	\\
        \N & 
	\N \ar[l]_{\one} \ar[r]^{!} \xtwocell[1,1]{}\omit{^<6>{\alpha_\one}} & 
	\onee &
        \N \ar[l]_{!} \ar[r]^{\one}  
	\xtwocell[1,-1]{}\omit{<-6>{\beta_{n}}}&
        \N 
	\\
	\N \ar[u]^{\one} \ar[d]_{\one}	& 
	& \onee \ar[u] \ar[d]& & \N \ar[u]_{\one} 
	\ar[d]^{\lambda z:nat. z*z} \\
	\N & 
	\N \ar[l]_{\one} \ar[r]^{!}  & 
	\onee &
        \N \ar[l]_{!} \ar[r]^{\one}  &
        \N 
}
$
\end{center}

{\noindent}with $\alpha_v = \langle \tmop{id}, \lambda x : \tmop{nat} . x + x
\rangle$ and $\beta_m = \langle \pi_1, \pi_2 \rangle = \tmop{id}$. The
pullback object is

\begin{center}
  $\xymatrix@=5mm{
\N^2 \\
\N^2 \ar[u]_{\one} \ar[d]^{<(\lambda x:nat.5) \: \circ \: \pi_1,\pi_2>} \\
\N^2 \\
\{(a,(b,a+a))\;|\;a,b\in\N\} 
\ar[u]_{<\pi_1,\pi_1 \: \circ \: \pi_2>} 
\ar[d]^{<\pi_1,\pi_2 \: \circ \: \pi_2>} \\
\N^2 \\
\N^2 \ar[u]_{\one} \ar[d]^{<\pi_1,(\lambda z:nat. z*z) \: \circ \: \pi_2>} \\
\N^2
}$

\end{center}
\[  \]

\section{\label{sec-sim}Simulation}

In this section we define a basic notion of simulation of HDA's. We show that
it carries over to categorical transition systems {\tmem{via}}
categorification.

\subsection{Simulations of HDA's}

\begin{definition}
  \label{def-hdasim}Let $s : S \longrightarrowlim ! \Sigma$ and $t : T
  \longrightarrowlim ! \Sigma$ be pointed HDA'a with points $S_{\bullet} \in
  S_0$ respectively $T_{\bullet} \in T_0$. A simulation $s \nrightarrow t$ is
  a relation
  \[ r \subseteq S_0 \times T_0 \]
  such that
  \begin{enumerateroman}
    \item $S_{\bullet} ( r ) T_{\bullet}$ ;
    
    \item $\forall x \in S_0 . \forall x' \in T_0 . \forall n \in \mathbbm{N}
    . \forall k \in S_n .$
    \[ x ( r ) x' \wedge \tmop{dom}_n ( k ) = x \Rightarrow \exists k' \in
       T_n . \tmop{dom}_n ( k' ) = x' \wedge \tmop{cod}_n ( k ) ( r )
       \tmop{cod}_n ( k' ) \wedge s_n ( k ) = t_n ( k' ) . \]
  \end{enumerateroman}
\end{definition}

The notion of simulation of definition \ref{def-hdasim} extrapolates the
notion of strong simulation of transition systems. It  admits a
characterization in terms of an appropriate notion of open maps.

\begin{definition}
  \label{def:hd-paths}Let $\left( S, < \right)$ be a strict total order. Given
  $x, y \in S$ let
  \[ x \prec y \Longleftrightarrowlim^{\tmop{def} .} x < y \hspace{0.5em}
     \wedge \hspace{0.5em} \not{\exists} z \in S. \hspace{0.25em} x < z < y \]
\end{definition}

\begin{definition}
  Let $\tmmathbf{\tmop{pcSet}}$ be the category of pointed cubical sets. Let
  $\mathbb{P} \subseteq \underset{}{\tmmathbf{\tmop{pcSet}}}$ be the full
  subcategory with objects $P \in \mathbb{P}$ such that
  \begin{enumerate}
    \item there is a subset $S \subseteq P_0$ which is a finite strict total
    order;
    
    \item the point is given by the smallest element of $S$;
    
    \item given $x, y \in S$
    \begin{enumerate}
      \item $x \nprec y \hspace{0.5em} \Rightarrow \hspace{0.5em} \forall n
      \in \mathbb{N} . \hspace{0.25em} P_n \left( x, y \right) = \emptyset$ ;
      
      \item given $x \prec y$ there is $n_{x, y} \in \mathbb{N}$ such that
      there is precisely one $T_{x, y} \in P_{n_{x, y}} \left( x, y \right)$.
      This $T_{x, y}$ is not a face and $P_n \left( x, y \right) = \emptyset$
      for $n < n_{x, y}$;
      
      \item given $z \in S$ and $x \prec y \prec z$, $T_{x, y}$ and $T_{y, z}$
      have $y$ as the only common face;
    \end{enumerate}
    \item any other non-degenerate $n$-cube in $P$ is a face of precisely one
    $T_{x, y}$.
  \end{enumerate}
  Objects of $\mathbb{P}$ are called \textit{paths}.
\end{definition}

Definition \ref{def:hd-paths} states that paths are series of $n$-transitions
glued at the endpoints of their main diagonals as for instance in

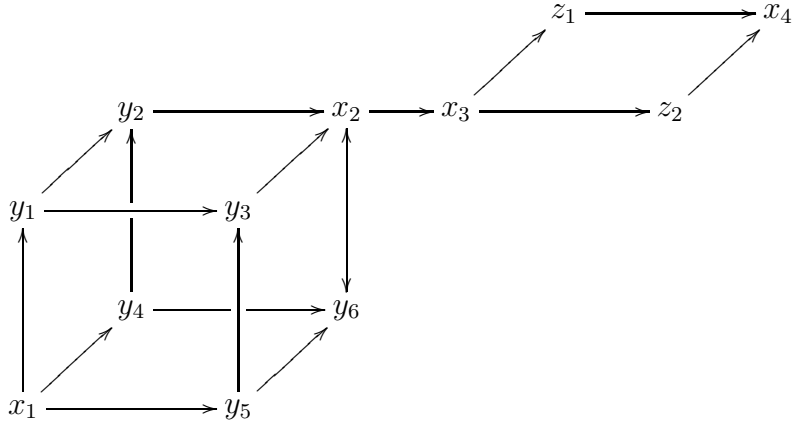
\begin{center}
  \xymatrix{
 & & & & & z_1 \ar[rr] & & x_4\\
 & y_2 \ar[rr] 
 & & x_2 \ar[dd] \ar[r] & x_3 \ar[rr] \ar[ur] & & z_2 \ar[ur] &
 \\ 
  y_1 \ar[ur] \ar[rr] 
  & & y_3 \ar[ur] & & & & & &
 \\ 
  & y_4 \ar'[r][rr] \ar'[u][uu]
   & & y_6 \ar[uu] & & & &
 \\ 
  x_1 \ar[rr]\ar[ur] \ar[uu]
  & & y_5 \ar[ur] \ar[uu] & & & & &
}


\end{center}

{\noindent}where $S = \left\{ x_1 \prec x_2 \prec x_3 \prec x_4 \right\}$. It
is easily seen from this example that $S$ induces a partition on $K_0
\setminus S$.

\begin{definition}
  A morphism $h : \hspace{0.25em} M \rightarrow K$ of pointed cubical sets is
  open provided $h_0$ is surjective and $h$ has the right lifting property
  with respect to $\mathbbm{P}_1$.
\end{definition}

\begin{lemma}
  \label{lem-zigzag}Let $\text{$h : \hspace{0.25em} M \rightarrow K$}$ be a
  morphism of pointed cubical sets such that $h_0$ is surjective. The
  following are equivalent.
  \begin{enumerate}
    \item $h$ is open ;
    
    \item $\forall n \in \mathbb{N} . \forall b \in K_n . \hspace{0.25em}
    \tmop{dom}_n ( b ) = h_0 ( x ) \Rightarrow \exists a \in M_n .
    \hspace{0.25em} h_n \left( a \right) = b$ .
  \end{enumerate}
\end{lemma}

\begin{theproof}
  Obvious.
\end{theproof}

\begin{proposition}
  Let $s : S \longrightarrowlim ! \Sigma$ and $t : T \longrightarrowlim !
  \Sigma$ be pointed HDA'a. The following are equivalent.
  \begin{enumerate}
    \item there is a simulation $s \nrightarrow t$;
    
    \item there is a span

    \begin{center}
      $
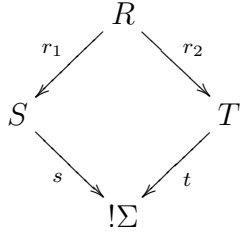
\xymatrix{ 
& R \ar[dr]^{r_2} \ar[dl]_{r_1} & \\
S \ar[dr]_{s} & & T \ar[dl]^{t} \\
& !\Sigma &
}$

    \end{center}

    {\noindent}in $\tmmathbf{\tmop{pcSet}} \hspace{0.25em} / \hspace{0.25em} !
    \Sigma$ with $r_1$ open.
  \end{enumerate}
\end{proposition}

\begin{theproof}
  Lemma \ref{lem-zigzag} allows the usual ''lift and project'' argument for
  any dimension.
\end{theproof}

A simulation of HDA'a can be characterized as a right homotopy with respect to
a certain model structure on $\tmmathbf{\tmop{pc} 2-\tmop{Cat}}$, the category
of pointed cubical 2-categories {\cite{getco}}.

\subsection{Simulations of Categorical Transition Systems}

\begin{definition}
  A functor is $\mathcal{C} ( \mathbbm{P} )$-{\tmem{open}} if it has the
  right lifting property with respect to any morphism in $\mathcal{C} (
  \mathbbm{P} )$.
\end{definition}

\begin{lemma}
  \label{lem-openmaps} The following are equivalent:
  \begin{enumerateroman}
    \item functor $\mathbbm{B} \longrightarrowlim^F \mathbbm{C}$ is
    $\mathcal{C} ( \mathbbm{P} )$-open;
    
    \item any $\mathbbm{C} \ni : \hspace{0.25em} F \left( B \right)
    \rightarrow C$ lifts under $F$.
  \end{enumerateroman}
\end{lemma}

\begin{theproof}
  Let $\tmmathbf{\tmop{Ord}} \subseteq \tmmathbf{\tmop{Cat}}$ be the full
  subcategory with finite ordinals as objects. Then
  \[ \mathcal{C} ( \mathbbm{P} ) \cong \tmmathbf{\tmop{Ord}} \]
  and the assertion follows by the usual ``lift and project'' argument.
\end{theproof}

\begin{definition}
  \label{def-sim}Suppose $G$ and $H$ are pointed reflexive graphs with points
  $G_{\bullet} \in G_0$ respectively $H_{\bullet} \in H_0$. Let $s :
  \mathcal{F} ( G ) \rightarrow \mathbf{\tmop{Span}}$ and $t : \mathcal{F (} H
  ) \rightarrow \mathbf{\tmop{Span}}$ be pointed categorical transition
  systems. A \textit{path simulation} from $s$ to $t$ is a relation $r
  \subseteq G_0 \times H_0$ such that
  \begin{enumerate}
    \item $G_{\bullet}  \hspace{0.25em} \left( r \right) H_{\bullet}$ ;
    
    \item $x \hspace{0.25em} \left( r \right) \hspace{0.25em} x'
    \hspace{0.5em} \hspace{0.5em} \& \hspace{0.5em} \hspace{0.5em} x
    \longrightarrowlim^f y \in \mathcal{F (} G ) \hspace{0.75em} \Rightarrow
    \hspace{0.75em} \exists \hspace{0.25em} x' \longrightarrowlim^{f'} y' \in
    \mathcal{F} ( H ) . \hspace{0.25em} y \hspace{0.25em} \left( r \right)
    \hspace{0.25em} y' \quad \& \quad s \left( f \right) = t \left( f'
    \right)$  .
  \end{enumerate}
\end{definition}

\begin{proposition}
  \label{prop-ctsopenmaps}Let $s : \hspace{0.25em} \mathcal{F} ( G )
  \rightarrow \mathbf{\tmop{Span}}$ and $t : \hspace{0.25em} \mathcal{F (} H )
  \rightarrow \mathbf{\tmop{Span}}$ be pointed categorical transition systems.
  The following are equivalent:
  \begin{enumerateroman}
    \item there is a path simulation from $s$ to $t$  ;
    
    \item there is a graph $R,$ a surjective graph homomorphism $d :
    \hspace{0.25em} R \rightarrow G$ and a functor $C : \hspace{0.25em}
    \mathcal{F} \left( R \right) \rightarrow \mathcal{F} \left( H \right)$
    such that

    \begin{center}
      $\xymatrix{
& \mathcal{F}R \ar[dl]_{\mathcal{F}d} \ar[dr]^C & \\
\mathcal{F}G \ar[dr]_s	& & \mathcal{F}H\ar[dl]^t \\
&\mathbf{Span} &
}$

    \end{center}

    {\noindent}commutes and $\mathcal{F} d$ is $\mathcal{C} ( \mathbbm{P}
    )$-open.
  \end{enumerateroman}
  
\end{proposition}

\begin{theproof}
  Lemma \ref{lem-openmaps} allows the usual ''lift and project'' argument for
  any dimension.
\end{theproof}

Proposition \ref{prop-ctsopenmaps} shows that path simulation is
{\tmem{formally}} a strong one. However, categorification brings free
categories over the control graphs into the game so we have composition of
individual transitions. In a sense, this blurs the difference between the
strong and the weak flavor of simulation. Consider for instance the
{\tmname{CIP}} programs
\[ \mathtt{x : \tmop{nat}} \hspace{0.5em} \hspace{0.5em} \vdash
   \hspace{0.5em} \mathtt{x : = 7} \]
and
\[ \mathtt{x : \tmop{nat}} \hspace{0.5em} \hspace{0.5em} \vdash \hspace{0.5em}
   \mathtt{x : = 5 ; x : = x + 2} \]
There is  the obvious path simulation
\[ \underline{\pi_{\mathtt{x \assign 7}}} \nrightarrow
   \underline{\pi_{\mathtt{x \assign 5 ; x \assign x + 2}}} \]
while there would be no simulation at all if we had more conservatively
rephrased strong simulation in terms of individual transitions only. As
illustrated by this example, the notion of path simulation is appropriate for
the study of refinement of specification.

However, the above path simulation is not a {\tmem{path bisimulation}}.
Indeed, in a path bisimulation, individual transitions need to be matched. The
notion of path bisimulation is thus quite crude, as crude as the traditional
notion of bisimulation. It is nonetheless a useful notion, yet one ingredient
is still missing. Consider for instance the {\tmname{CIP}} programs
\[ \mathtt{x : \tmop{nat}} \;|\; \mathtt{p : \tmop{nat}} \hspace{0.5em}
   \hspace{0.5em} \vdash \hspace{0.5em} \mathtt{\tmop{nop}} \]
and
\[ \mathtt{x : \tmop{nat}} \hspace{0.5em} \;|\; \mathtt{p : \tmop{nat}}
   \hspace{0.5em} \vdash \hspace{0.5em} \mathtt{p! (2 \ast x)} \]
There is the nonsense path bisimulation
\[ r : \underline{\pi_{\mathtt{\tmop{nop}}}} \nrightarrow
   \underline{\pi_{\mathtt{p! (2 * x)}}} \]
since we have 
\[
C_{\mathtt{nop}} \cong C_{\mathtt{p!(2 * x)}} \cong 
J
\]
with $J$ the interval category $- \longrightarrowlim^t +$ and
\[
\underline{\pi_{\mathtt{nop}}} \cong 
\underline{\pi_{\mathtt{p!(2 * x)}}}
\]
Indeed, it is necessary 
to \textit{classify} the transitions in addition to
the computations they carry. Since we got rid of the ``syntactic'' labels, a
different classifying device is required. The latter is given by the map to
the interface. This leads to considering bisimulation in the bicategory
$\text{$\tmmathbf{\tmop{Span}} ( \mathcal{F} /\!/
\tmmathbf{\tmop{Span}} ( \tmmathbf{\tmop{Set}}$))}$ as in {\cite{conpca}}.
Roughly, the semantics of $\mathtt{x : \tmop{nat}} \;|\;\mathtt{p :
\tmop{nat}} \;
\vdash \; \mathtt{\tmop{nop}}$ is then the span of
categorical transition systems
\[ \iota \leftarrow \underline{\pi_{\mathtt{\tmop{nop}}}} \rightarrow \iota \]
with $\iota$ the free category with one object $\star$ and one generator $p$.
Both legs of the span are the identity lax (representable) transformation.
On the other
hand, the semantics of $\mathtt{x : \tmop{nat}} \; |\;\mathtt{p :
\tmop{nat}} \;\vdash \; \mathtt{p! (2 * x)}$ is the
span of categorical transition systems
\[ \iota \leftarrow \underline{\pi_{\mathtt{p! (2 * x)}}} \rightarrow \iota
\]
The left leg is the identity lax transformation while the
component of the right one at $t$ is the morphism of spans
\[
\langle id,\lambda x \in \mathbb{N}.\: 2*x \rangle :\;
(id,\mathbb{N},!)
\longrightarrow
(\pi_1,\mathbb{N} \times \mathbb{N},!)
\]
(c.f. section \ref{sec-2-categorific}). Hence, the span
\[ \underline{\pi_{\mathtt{\tmop{nop}}}} \leftarrow \rho \rightarrow
   \underline{\pi_{\mathtt{p! (2 * x)}}} \]
of $\text{$\mathcal{C} ( \mathbbm{P} )$-open}$ maps witnessing the nonsense
path bisimulation $r$ does NOT make the diagram

\begin{center}
  $\xymatrix{ 
& \underline{\pi_{\mathtt{nop}}} \ar[dl] \ar[dr] &
\\
\iota & \rho \ar[u] \ar[d] & \iota
\\
& \underline{\pi_{\mathtt{p ! (2*x)}}} \ar[ul] \ar[ur] &
}$

\end{center}

{\noindent}commute so there is no bisimulation in this setting. Since
$\mathcal{C} ( \mathbbm{P} )$-open maps form a {\tmem{cover system}}, i.e.
isomorphisms are $\mathcal{C} ( \mathbbm{P} )$-open maps and $\mathcal{C} (
\mathbbm{P} )$-open maps are composition- as well as pullback-stable, the
construction of {\tmem{process categories}} {\cite{conpca}} applies.

\section{\label{sec-conc}Concluding Remarks}

We introduced the 1-coskeletal synchronization and showed how it can be
applied to the study of message-passing at hand of a realistic
application. We have good reasons to believe that other synchronization 
paradigms e.g. semaphores may turn out to
be expressible this way. The present paper provides a thorough harnessing of the
semantics of {\tmname{CIP}}-like message-passing programming languages. 
What makes the strength of the approach is that the semantics in question
is given in a direct way, i.e. without coding an imperative language into
some process calculus.
These principles have
actually been used in the design and implementation of a deductive
model-checking tool {\cite{memo-book}}, yet the potential has only been
scratched on the surface. What remains to do is to set up a relevant modal
logic.

\bibliographystyle{abbrv}
\bibliography{street,benabou,cat,johnstone,goub,serre,duskin,winskel,rbrown,grandis,kan,lindsay,me,kw,koslo,cockett,jacobs}

\begin{thebibliography}{10}

\bibitem{bena}
J.~B\'{e}nabou.
\newblock {\em Introduction to Bicategories}, volume~47 of {\em Lecture Notes
  in Mathematics}.
\newblock Springer, 1967.

\bibitem{bor1}
F.~Borceux.
\newblock {\em Handbook of Categorical Algebra 1}.
\newblock Cambridge University Press, 1994.

\bibitem{conpca}
J.~R.~B. Cockett and D.~A. Spooner.
\newblock Constructing process categories.
\newblock {\em Theoretical Computer Science}, 177(1), April 1997.

\bibitem{duskin-simtriple}
J.~Duskin.
\newblock {\em Simplicial methods and the interpretation of "triple"
  cohomology}, volume 163 of {\em Memoirs of the AMS}.
\newblock American Mathematical Society, 1975.

\bibitem{lindspap}
L.~Erington.
\newblock On the semantics of message passing processes.
\newblock In {\em Proceedings of CTCS99}, 1999.

\bibitem{linds}
L.~Erington.
\newblock {\em Twisted Systems}.
\newblock PhD thesis, Department of Computing, Imperial College, London, 1999.

\bibitem{eric:cmcim}
E.~Goubault.
\newblock Labelled cubical sets and asynchronous transitions systems: an
  adjunction.
\newblock In A.~Kurz, editor, {\em Electronic Notes in Theoretical Computer
  Science}, volume~68. Elsevier Science Publishers, 2002.

\bibitem{hha}
E.~Goubault.
\newblock Some geometric perspectives in concurrency theory.
\newblock {\em Homology, Homotopy and Applications}, 2002.

\bibitem{gr-mau}
M.~Grandis and L.~Mauri.
\newblock Cubical sets and their site.
\newblock 8:185--211, 2003.

\bibitem{bartbook}
B.~Jacobs.
\newblock {\em {C}ategorical {L}ogic and {T}ype {T}heory}.
\newblock Number 141 in Studies in Logic and the Foundations of Mathematics.
  North Holland, 1999.

\bibitem{JohnstonePT:dicf}
P.~T. Johnstone.
\newblock A note on discrete conduch\'e fibrations.
\newblock {\em Theory and Applications of Categories}, 5(1):1--11, 1999.

\bibitem{kan1}
D.~Kan.
\newblock On c.s.s. complexes.
\newblock {\em American Journal of Mathematics}, 79:449--476, 1957.

\bibitem{getco}
K.Hess, P.E.Parent, A.Tonks, and K.Worytkiewicz.
\newblock Simulations as homotopies.
\newblock {\em Electronic Notes in Theoretical Computer Science}, 90(5), 2004.
\newblock Forthcoming.

\bibitem{koslo}
J.~Koslowski.
\newblock Categorical transition systems and the comprehension scheme.
\newblock Talk at PSSL 80, Cambridge, April 2004.

\bibitem{alg-cub}
R.Brown and P.J.Higgins.
\newblock On the algebra of cubes.
\newblock {\em JPAA}, 21:233--360, 1981.

\bibitem{serre-phd}
J.-P. Serre.
\newblock {\em Homologie Singuli{\`e}re des Espaces Fibr{\'e}s. Applications.}
\newblock PhD thesis, Ecole Normale Sup{\'e}rieure, 1951.

\bibitem{memo-book}
C.~Sprenger and K.~Worytkiewicz.
\newblock {\em Formal methods and models for system design}, chapter A
  verification methodology for infinite-state message-passing systems.
\newblock Kluwer, 2004.
\newblock Forthcoming.

\bibitem{street-cat-struct}
R.~Street.
\newblock {\em Handbook of Algebra}, chapter Categorical Structures, pages
  529--577.
\newblock Elsevier Science, 1996.

\bibitem{NielsenM:modc1}
G.~Winskel and M.~Nielsen.
\newblock Models for concurrency.
\newblock In S.~Abramsky, D.~Gabbay, and T.~S.~E. Maibaum, editors, {\em
  Handbook of Logic in Computer Science}. Oxford University Press, 1995.

\bibitem{me}
K.~Worytkiewicz.
\newblock {\em Components and Synchronous Communication in Categories of
  Processes}.
\newblock PhD thesis, Swiss Federal Institute of Technology, Lausanne, 2000.

\bibitem{kw:cmcim}
K.~Worytkiewicz.
\newblock Concrete process categories.
\newblock In A.~Kurz, editor, {\em Electronic Notes in Theoretical Computer
  Science}, volume~1. Elsevier Science Publishers, 2002.

\bibitem{kw:ctcs}
K.~Worytkiewicz.
\newblock Paths and simulations.
\newblock In R.~Blute and P.~Selinger, editors, {\em Electronic Notes in
  Theoretical Computer Science}, volume~69. Elsevier, 2003.

\end{thebibliography}

\end{document}